\documentclass[%
 reprint,
 superscriptaddress,
 amsmath,amssymb,
 aps,
 floatfix,
]{revtex4-2}

\usepackage{comment}
\usepackage{eqnarray}
\usepackage{physics}
\usepackage{dsfont}
\usepackage{tensor}
\usepackage{amsthm}
\usepackage{amsmath}
\usepackage{bbold}
\usepackage{fancyhdr}
\usepackage{subfigure}
\usepackage{epsfig}
\usepackage{bbm}
\usepackage{bm}
\usepackage{tikz}
\usepackage{float}
\usepackage{xcolor}
\usepackage{amsmath}
\usepackage{graphicx}
\usepackage{dcolumn}
\usepackage{bm,braket}
\usepackage{comment}

\usepackage{soul}

\renewcommand{\selectlanguage}[1]{}
\usepackage{hyperref}

\begin{document}

\preprint{}

%\title{Chirality Controlled Accumulation of Active Brownian Spinners in Activity Gradients}
\title{Active Transport of Cargo-Carrying and Interconnected Chiral Particles}

\author{Bhavesh Valecha}
\affiliation{Mathematisch-Naturwissenschaftlich-Technische Fakult\"at, Institut f\"ur Physik, Universit\"at Augsburg, Universit\"atsstra{\ss}e 1, 86159 Augsburg, Germany}
\author{Hossein Vahid}
\affiliation{Leibniz-Institut f\"ur Polymerforschung Dresden, Institut Theory der Polymere, 01069 Dresden, Germany}
\author{Pietro Luigi Muzzeddu}
\affiliation{University of Geneva, Department of Biochemistry, CH-1205 Geneva, Switzerland}
\author{Jens-Uwe Sommer}
\affiliation{Leibniz-Institut f\"ur Polymerforschung Dresden, Institut Theory der Polymere, 01069 Dresden, Germany}
\affiliation{Technische Universit\"at Dresden, Institut f\"ur Theoretische Physik, 01069 Dresden, Germany}
\author{Abhinav Sharma}
\affiliation{Mathematisch-Naturwissenschaftlich-Technische Fakult\"at, Institut f\"ur Physik, Universit\"at Augsburg, Universit\"atsstra{\ss}e 1, 86159 Augsburg, Germany}
\affiliation{Leibniz-Institut f\"ur Polymerforschung Dresden, Institut Theory der Polymere, 01069 Dresden, Germany}

\date{\today}

\begin{abstract}
Directed motion up a concentration gradient is crucial for the survival and maintenance of numerous biological systems, such as sperms moving towards an egg during fertilization or ciliates moving towards a food source. In these systems, chirality---manifested as a rotational torque---plays a vital role in facilitating directed motion. While systematic studies of active molecules in activity gradients exist, the effect of chirality remains little studied. In this study, we examine the simplest case of a chiral active particle connected to a passive particle in a spatially varying activity field. We demonstrate that this minimal setup can exhibit rich emergent tactic behaviors, with the chiral torque serving as the tuning parameter. Notably, when the chiral torque is sufficiently large, even a small passive particle enables the system to display the desired accumulation behavior. Our results further show that in the dilute limit, this desired accumulation behavior persists despite the presence of excluded volume effects. Additionally, interconnected chiral active particles exhibit emergent chemotaxis beyond a critical chain length, with trimers and longer chains exhibiting strong accumulation at sufficiently high chiral torques. This study provides valuable insights into the design principles of hybrid bio-molecular devices of the future.

\end{abstract}

\maketitle

\section{Introduction}

Active matter systems represent a broad class of nonequilibrium systems that convert environmental energy into directed motion, enabling them to respond to external stimuli~\cite{hanggi_artificial_2009, ramaswamy_mechanics_2010, marchetti_hydrodynamics_2013, bechinger_active_2016, ramaswamy_active_2017}. One of the most well-known forms of directed transport in such systems is chemotaxis, the ability of microorganisms like \textit{Escherichia coli} to navigate chemical gradients~\cite{berg_e_2004}. Bacteria achieve this via a run-and-tumble strategy, adjusting their motion in response to nutrient gradients~\cite{cates_diffusive_2012, kromer_chemokinetic_2020}. Synthetic analogs of chemotaxis, such as phoretic Janus colloids with catalytic coatings, have been experimentally realized, demonstrating self-propulsion in response to chemical gradients~\cite{marchetti_hydrodynamics_2013, bechinger_active_2016}. 

To model chemotaxis theoretically, various approaches have been developed. Simple active Brownian particle (ABP) models~\cite{howse_self-motile_2007} capture self-propulsion and persistence but typically lead to accumulation in low-fuel regions due to orthokinesis~\cite{fraenkel_orientation_1941, schnitzer_theory_1993, sharma_brownian_2017, caprini_dynamics_2022}. To overcome this limitation, explicit coupling to concentration gradients~\cite{keller_model_1971, liebchen_synthetic_2018, stark_artificial_2018}, feedback mechanisms~\cite{qian_harnessing_2013, mano_optimal_2017, massana-cid_rectification_2022}, and external control via light or magnetic fields~\cite{lozano_phototaxis_2016, lozano_diffusing_2019, martinez-pedrero_magnetic_2015} have been proposed. It has also been demonstrated that the presence of torques aligning the self-propulsion of active particles, either parallel or anti-parallel to activity gradients, can suppress or enhance their tactic behavior, respectively~\cite{geiseler_self-polarizing_2017,geiseler_chemotaxis_2016}. More recently, alternative strategies have been devised, in which directed motion arises as an emergent property without external feedback. Examples include coupling an active particle to a passive particle~\cite{vuijk_chemotaxis_2021}, active colloidal molecules with fixed orientations~\cite{vuijk_active_2022}, and chains of active particles~\cite{muzzeddu_migration_2024}.\\

A relatively underexplored aspect of chemotactic transport is the role of chirality, which is ubiquitous in both biological locomotion and synthetic active systems. Chirality introduces an intrinsic rotational component to active motion, fundamentally altering transport properties~\cite{liebchen_chiral_2022, lowen_chirality_2016}. Many biological microswimmers, such as bacteria~\cite{crenshaw_new_1996,lauga_swimming_2006}, sperm cells~\cite{riedel_self-organized_2005,friedrich_chemotaxis_2007}, and flagellated algae~\cite{drescher_dancing_2009}, exhibit helical or rotational swimming due to internal asymmetries. Advances in fabrication techniques have enabled the design of artificial chiral active particles, including colloidal spinners, magnetic rotors, and chiral granular particles, revealing novel transport properties~\cite{kummel_circular_2013, matsunaga_controlling_2019, arora_emergent_2021, bililign_motile_2022}. Chirality has also been shown to suppress wall adhesion and give rise to surface currents~\cite{caprini_active_2019}, induce phase separation~\cite{lei_nonequilibrium_2019}, generate spontaneous angular momentum in active crystals~\cite{marconi_spontaneous_2025}, and break time-reversal symmetry leading to unidirectional wave propagation~\cite{ventejou_susceptibility_2021} and topologically protected excitations~\cite{souslov2017}.

In this work, we introduce chirality as a control parameter in active-passive transport. Our modeling closely follows the approach developed in our previous works \cite{vuijk_chemotaxis_2021, vuijk_active_2022, muzzeddu_migration_2024}. We explore the impact of chirality on the accumulation of a passive cargo attached to an active chiral particle within an activity gradient. Our results demonstrate that chirality fundamentally changes the transition to chemotaxis: even when the passive particle is small, a sufficiently large chiral rotational torque enables preferential accumulation in high-activity regions.
This is in contrast to previous findings, where a large passive particle was required to induce chemotaxis~\cite{vuijk_chemotaxis_2021}.
We further emphasize the robustness of the effect of chirality by performing simulations with excluded volume effects.

Additionally, chirality is naturally present in biological active systems composed of elongated structures, such as cytoskeletal filaments~\cite{dunajova2023} and bacterial protofilaments~\cite{shi2020}. 
We extend our analysis to chains of chiral active particles. We find that while monomers and dimers accumulate in the low-activity regions, trimers accumulate in the high-activity regions at high chirality levels. As chirality increases in longer chains, their accumulation in high-activity regions becomes more pronounced.

\section{The model}
\label{sec:the_model}
\begin{figure}[htp!]
    \centering
    \begin{tikzpicture}
    \node at (0,0) {\includegraphics[width=.5\textwidth]{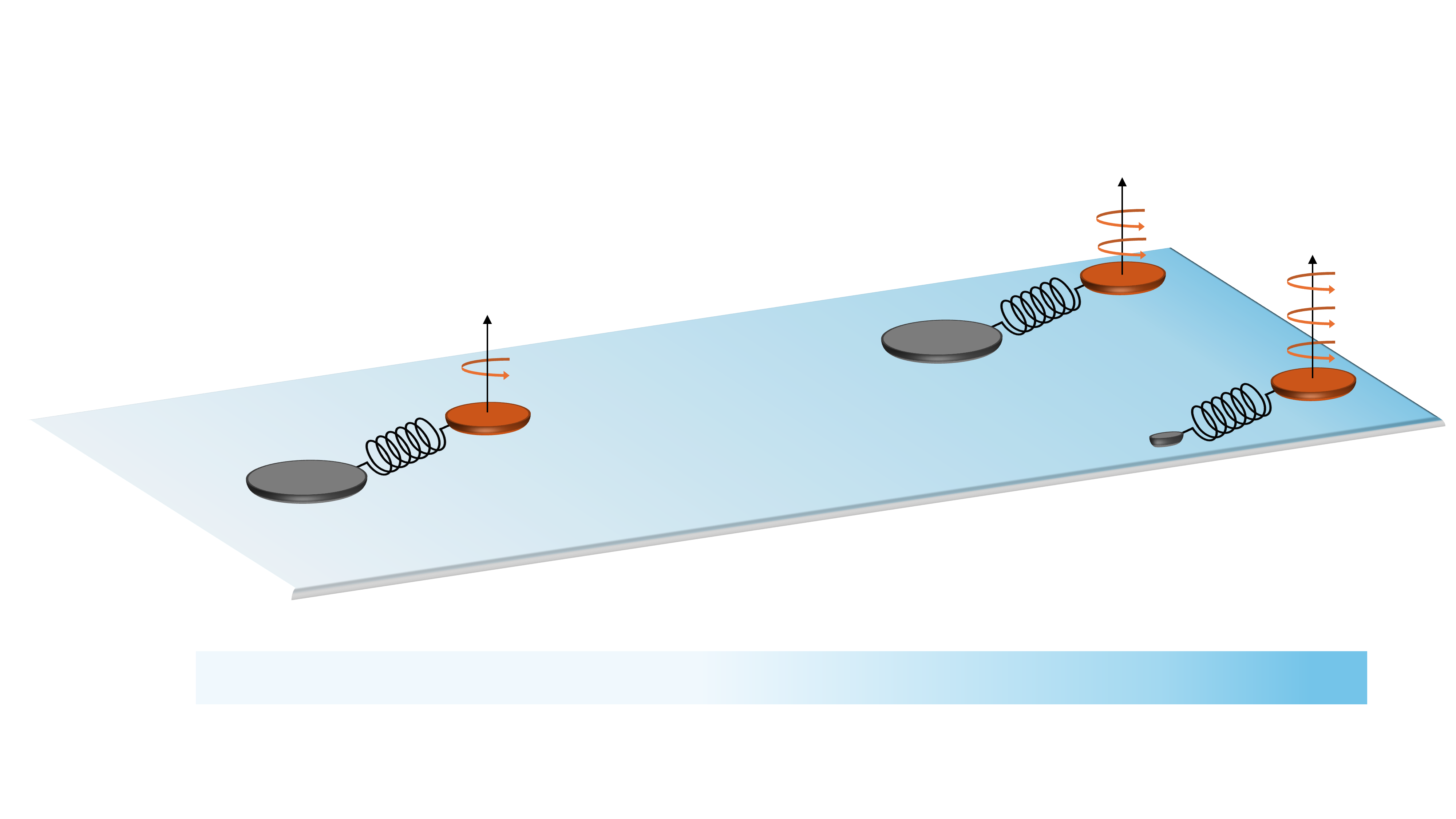}};
    \node at (.35,-2.1) {Activity gradient};
    \node at (-3.2,-.5) {\fontsize{7}{5}\selectfont (a)};
    \node at (.75,.3) {\fontsize{7}{5}\selectfont (b)};
    \node at (2.35,-.175) {\fontsize{7}{5}\selectfont (c)};
    \node at (-1.4,.75) {\fontsize{7}{5}\selectfont $\omega_{\text{a}}$};
    \node at (2.5,1.55) {\fontsize{7}{5}\selectfont $\omega_{\text{b}}$};
    \node at (3.65,1.1) {\fontsize{7}{5}\selectfont $\omega_{\text{c}}$};
    \end{tikzpicture}
    \caption{Schematic of composites of a chiral active particle (orange) connected to a passive particle (grey) via a harmonic spring in a spatially varying activity field. The color bar below the schematic indicates the gradient in the activity field increasing in magnitude from left to right. The number of curved arrows indicates the strength of the chiral torque on the active particles about an axis perpendicular to the plane. Here, $\omega_{\text{a}}<\omega_{\text{b}}<\omega_{\text{c}}$. Composites with a small torque on the active particle accumulate in low-activity regions (composite a), whereas those with a large chiral torque accumulate in high-activity regions (composite b). Furthermore, for a very large chiral torque on the active particle, a small passive particle is sufficient to facilitate accumulation in high-activity regions (composite c).}
\label{fig:sketch}
\end{figure}

We consider a two dimensional system of a chiral active particle and a passive particle interacting through a force $\boldsymbol{F}(\boldsymbol{r})$ in a spatially varying activity field (see Fig.~\ref{fig:sketch}). We model this active-passive composite with the overdamped Langevin dynamics given by the following coupled stochastic differential equations:
\begin{subequations}
\begin{align}
    \frac{d\boldsymbol{ r}_1}{dt} &= \frac{1}{\gamma}\boldsymbol{F}+\frac{1}{\gamma}f_s(\boldsymbol{r}_1)\boldsymbol{p}+\sqrt{\frac{2T}{\gamma}}\,\boldsymbol{\xi}_1(t),\\
    \frac{d\theta}{dt} &= \omega + \sqrt{2D_R}\,\eta(t),\\
    \frac{d\boldsymbol{ r}_2}{dt} &= -\frac{1}{q\gamma}\boldsymbol{F}+\sqrt{\frac{2T}{q\gamma}}\,\boldsymbol{\xi}_2(t).
\end{align}
\label{eq:chiral_active_passive_dimer_dynamics}
\end{subequations}   
Here, $\boldsymbol{r}_1$ and $\boldsymbol{r}_2$ are the coordinates of the active and passive particle, respectively. The friction coefficient of the passive particle is $q$ times that of the active particle, which is given by $\gamma$. For spherical active and passive particles in a Newtonian fluid, $q$ is the ratio of their radii (sizes). The interaction force between the active and passive particles is modeled as a harmonic spring force $\boldsymbol{F}=-k\,(r-l_0)\boldsymbol{\hat{r}}$, where $\boldsymbol{r}=\boldsymbol{r}_1-\boldsymbol{r}_2$, the spring constant is $k>0$ and the rest length is $l_0$. The self-propulsion force of the active particle, whose magnitude is given by $f_s(\boldsymbol{r}_1)$, is aligned with the rotational degree of freedom $\boldsymbol{p}=\big(\cos{\theta},\,\,\sin{\theta}\big)$, where $\theta$ is the polar angle measured with respect to the $x$-axis. $f_s$ will henceforth be referred to as the activity field. The active particle also experiences a constant chiral torque $\omega$, in the direction perpendicular to the plane of motion, as well as rotational diffusion characterized by the diffusion coefficient $D_R$. The temperature $T$ of the thermal bath is measured in units such that the Boltzmann constant $k_B$ is set to unity. The stochastic terms $\boldsymbol{\xi}_1$, $\boldsymbol{\xi}_2$ and $\eta$ are zero-mean independent Gaussian white noises, with correlations:
\begin{equation}
\begin{split}
\langle\boldsymbol{\xi}_i(t)\otimes \boldsymbol{\xi}_i(t')\rangle &= \boldsymbol{\mathds{1}}\delta (t-t')\,,\\
\langle\eta(t)\eta(t')\rangle &= \delta (t-t')\,.
    \end{split}
    \label{eq:noise_correlations}
\end{equation}
with $\otimes$ denoting the outer product and $i \in \{1,2\}$. 

The dynamics of the active-passive composite can be equivalently described by the collective coordinates, which we identify as the center of friction $\bm{R}$ and the bond coordinates $\bm{r}$, defined as:
\begin{equation}
\begin{split}
    \boldsymbol{R}&=\frac{1}{1+q}\boldsymbol{r}_1+\frac{q}{1+q}\boldsymbol{r}_2,\\    
    \boldsymbol{r}&=\boldsymbol{r}_1-\boldsymbol{r}_2.
\end{split}
\label{eq:coordinate_transformation}
\end{equation}
To analyze the spatial regions where the active-passive composite preferentially accumulates in the long-time limit, it is convenient to express the evolution equation for the one-time joint probability density $P(\boldsymbol{R},\boldsymbol{r},\theta,t)\equiv P$. Since the stochastic dynamics in Eq.~\eqref{eq:chiral_active_passive_dimer_dynamics} is Markovian, this is simply given by the Fokker-Planck equation~\cite{risken_fokker-planck_1996}:
\begin{equation}
\begin{split}
    \dfrac{\partial}{\partial t}\,P =& -\nabla_{\boldsymbol{R}}\cdot\bigg[\dfrac{1}{1+q}\dfrac{1}{\gamma}f_{s}\boldsymbol{p}P - \dfrac{1}{1+q}\dfrac{T}{\gamma}\nabla_{\boldsymbol{R}}P\bigg] \\
    &-\nabla_{\boldsymbol{r}}\cdot\bigg[\dfrac{1+q}{q}\dfrac{1}{\gamma}\boldsymbol{F}P + \dfrac{1}{\gamma}f_{s}\boldsymbol{p}P -\dfrac{1+q}{q}\dfrac{T}{\gamma}\nabla_{\boldsymbol{r}}P\bigg]\\
    &-\omega\partial_{\theta}P+ D_R\partial_{\theta}^{2}P.   
\end{split}
\label{eq:fokker_planck}
\end{equation}
Here, the symbol $\cdot$ represents a single contraction, $\nabla_{\boldsymbol{R}}$ and $\nabla_{\boldsymbol{r}}$ represent derivatives with respect to $\boldsymbol{R}$ and $\boldsymbol{r}$, and $ \partial_{\theta}$ is the rotation operator in two dimensions.

We are interested in the spatial regions where the active-passive composite preferentially accumulates in the steady state. To this end, we attempt a coarse-grained description of our system at the mean-field level akin to the analysis in Refs.~\cite{vuijk_chemotaxis_2021,muzzeddu_taxis_2023,muzzeddu_active_2022,cates_when_2013,solon_active_2015,adeleke-larodo_non-equilibrium_2020}, where we use the center of friction as a proxy for the position of the composite. This interpretation holds whenever the typical distance between the two particles is sufficiently small compared to the characteristic length scale of variation of the activity field. This gives us access to physically relevant fields such as the position density, orientation, etc. Particularly, we begin by performing a Cartesian multipole expansion of the probability density $P$ in the eigenfunctions of the operator $\partial_{\theta}^{2}$~\cite{muzzeddu_active_2022}:
\begin{equation}
  P(\boldsymbol{R},\boldsymbol{r},\theta,t) = \phi + \boldsymbol{\sigma}\cdot\boldsymbol{p} + \boldsymbol{\mu}\colon\big(\boldsymbol{p}\boldsymbol{p}-\mathds{1}/2\big) + \boldsymbol{\Theta}(P),
\label{eq:multipole_expansion}
\end{equation} 
where $\phi, \boldsymbol{\sigma}$ and $\boldsymbol{\mu}$ are functions of $\boldsymbol{R}$ and $\boldsymbol{r}$, and $\colon$ denotes a double contraction. Specifically, these correspond to the positional probability density, the average polarization, and the nematic tensor, respectively. The term $\boldsymbol{\Theta}(P)$ contains the dependencies on all the higher-order modes. Projecting Eq.~\eqref{eq:fokker_planck} onto the respective eigenfunctions, we obtain a hierarchy of time-evolution equations for the modes ($\phi, \boldsymbol{\sigma}, \boldsymbol{\mu},\ldots$), details of which can be found in the Supplementary Material (SM). %~\cite{supplementary_material}.
Importantly, we note that the slowest mode of the dynamics is the positional probability density $
2\pi\phi(\boldsymbol{R,r},t) = \int \dd\theta P(\boldsymbol{R,r},\theta,t)$, obtained by integrating out the rotational degree of freedom $\bm{p}$. In fact, $\phi(\boldsymbol{R,r},t)$ is a conserved quantity and satisfies a continuity equation:
\begin{equation}
\begin{split}
    \frac{\partial \phi}{\partial t}\, =& -\nabla_{\boldsymbol{R}}\cdot\bigg[-\frac{T}{\gamma}\frac{1}{1+q}\nabla_{\boldsymbol{R}}\phi + \frac{1}{1+q}\frac{1}{2\gamma}f_s\boldsymbol{\sigma}\bigg]\\ 
    &-\nabla_{\boldsymbol{r}}\cdot\left[\frac{1}{2\gamma}f_s \boldsymbol{\sigma}+\frac{1+q}{q}\frac{1}{\gamma}\boldsymbol{F}\phi -\frac{T}{\gamma}\frac{1+q}{q}\nabla_{\boldsymbol{r}} \phi\right].
\end{split}
\label{eq:phi_dynamics_transformed}
\end{equation}
Additionally, the orientation field $\boldsymbol{\sigma}(\boldsymbol{R,r},t) = \int \dd\theta\, \boldsymbol{p}P(\boldsymbol{R,r},\theta,t)$, which is related to the conditional average of the polarization vector $\bm{p}$ at fixed position of the active-passive composite, evolves in times as:
\begin{equation}
\begin{split}
    \frac{\partial \boldsymbol{\sigma}}{\partial t} =& -\big(D_R\,\boldsymbol{\mathds{1}}-\omega\,\boldsymbol{\varepsilon}\big)\cdot\boldsymbol{\sigma}-\frac{1}{\gamma}\left(\frac{1}{1+q}\nabla_{\boldsymbol{R}} +\nabla_{\boldsymbol{r}} \right)(f_s\phi)\\
    &-\frac{1}{\gamma}\frac{1+q}{q}\nabla_{\boldsymbol{r}}\cdot(\boldsymbol{F}\boldsymbol{\sigma}) + \mathcal{O}(\nabla_{\boldsymbol{R}}^{2}),
\end{split}
\label{eq:sigma_dynamics_transformed}
\end{equation}
where the dependence on the higher-order modes is captured in $\mathcal{O}(\nabla_{\boldsymbol{R}}^{2})$. It is important to note that $\boldsymbol{\sigma}(\boldsymbol{R,r},t)$ is a fast mode due to the presence of the sink term $\big(D_R\,\boldsymbol{\mathds{1}}-\omega\,\boldsymbol{\varepsilon}\big)\cdot\boldsymbol{\sigma} = \boldsymbol{\mathds{L}}^{-1}\boldsymbol{\sigma}$, where $\boldsymbol{\varepsilon}$ is the \emph{Levi-Civita} tensor in two dimensions. The eigenvalues of the matrix $\boldsymbol{\mathds{L}}$ govern the decay-timescale of the orientation field $\boldsymbol{\sigma}$. This implies that $\boldsymbol{\sigma}$ can be approximated to be quasi-static at the timescale of variations in $\phi$. Furthermore, we assume that the activity field is slowly varying in space, i.e., the gradients of $f_{s}(\boldsymbol{R})$ are i) small compared to the persistence length of the chiral active particle, and ii) small compared to the separation between the active and passive particles. This assumption implies that $\phi(\boldsymbol{R},\boldsymbol{r},t)$ also has small spatial variations, as well as the contribution from the higher-order modes $\mathcal{O}(\nabla_{\boldsymbol{R}}^{2})$ to Eq.~\eqref{eq:sigma_dynamics_transformed} can be neglected. The details of the coarse-graining and the validity of the approximation can be found in the SM.
\begin{figure}[t]
\centering
  \includegraphics[width=9.25cm, height=7.25cm]{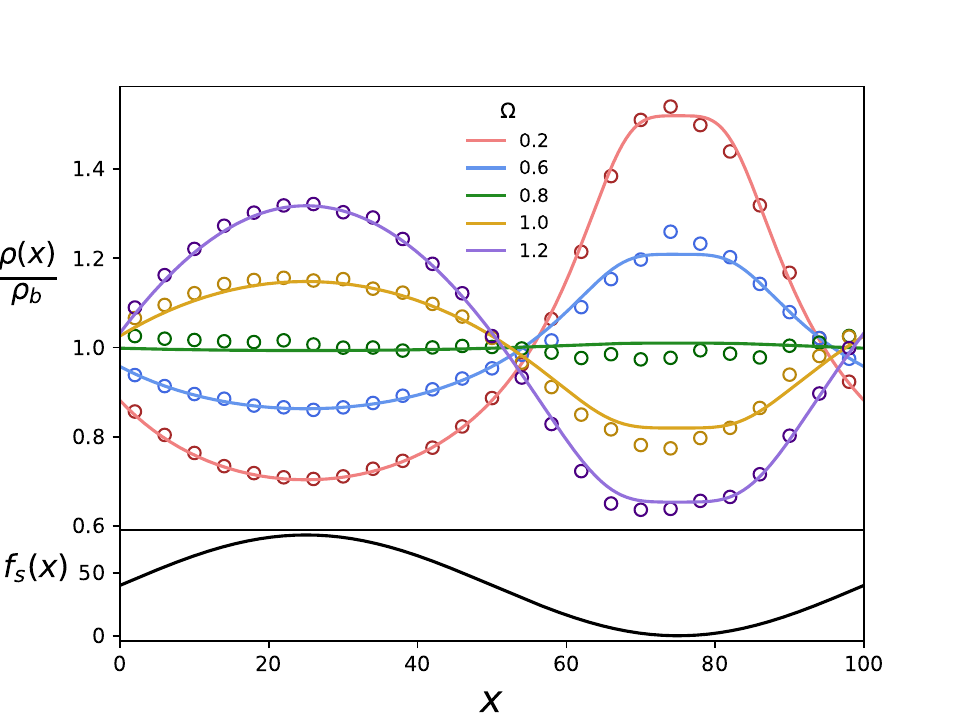}
  \caption{Steady state density distribution (top panel) of the active-passive composites connected via a spring with $l_0=0$ for different chiral torques on the chiral active particle. The solid lines are plotted using the theoretical result (Eq.~\eqref{eq:steady_state_density} and~\eqref{eq:epsilon_spring}) and the symbols are obtained from Langevin dynamics simulations of Eq.~\eqref{eq:chiral_active_passive_dimer_dynamics}. The accumulation behavior of the composite is controlled by the intensity of the chiral torque $\Omega$ for a fixed size of the passive particle. The composite accumulates in regions of high-activity for rapidly spinning active particles and in regions of low-activity for slowly spinning active particles, respectively. The bottom panel shows the activity field $f_s(x) = 40\big[1+\sin(2\pi x/L)\big]$ experienced by the chiral active particle. The $y$-axis in the top panel is normalized with the bulk density, defined as $\rho_b = 1/L$, where $L=100$ is the simulation box size with periodic boundary conditions. The parameters of the simulation are $k_{B}T=1.0$, $k=14.0$, $\gamma=1.0$, $D_{R} = 10.0$, $q = 2.0$, and the integration time step $\Delta t=D_R\times10^{-5}$.}
\label{fig:steady_state_density_spring_zero_restlenght}
\end{figure}  

\begin{figure*}[t]
  \centering
  \subfigure{\includegraphics[width=8.5cm,height=6.5cm]{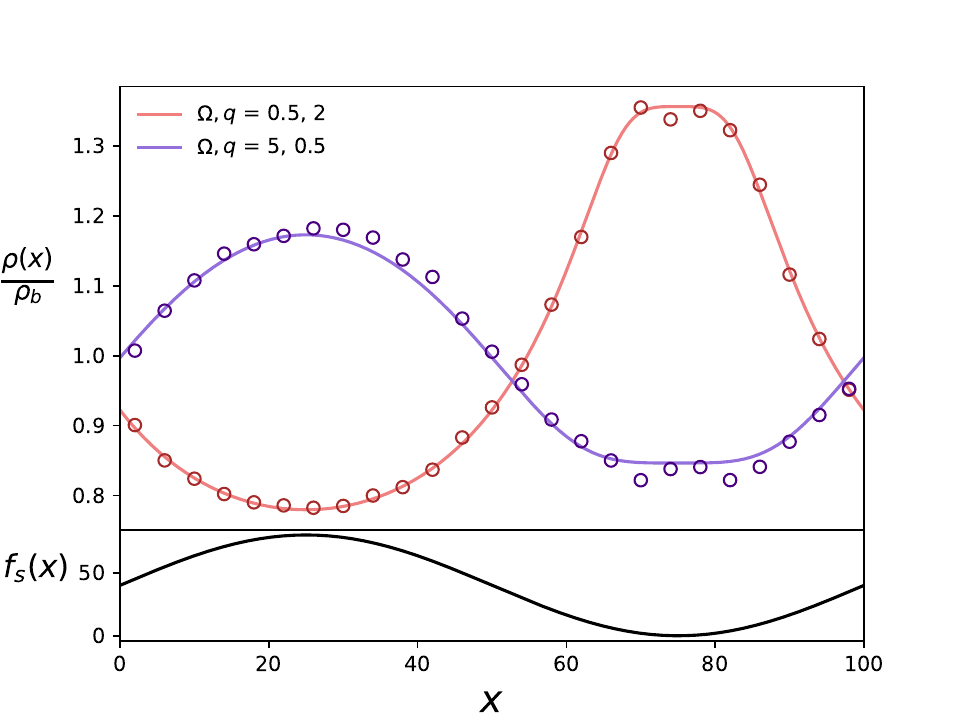}}\quad
  \subfigure{\includegraphics[width=8.5cm,height=6.5cm]{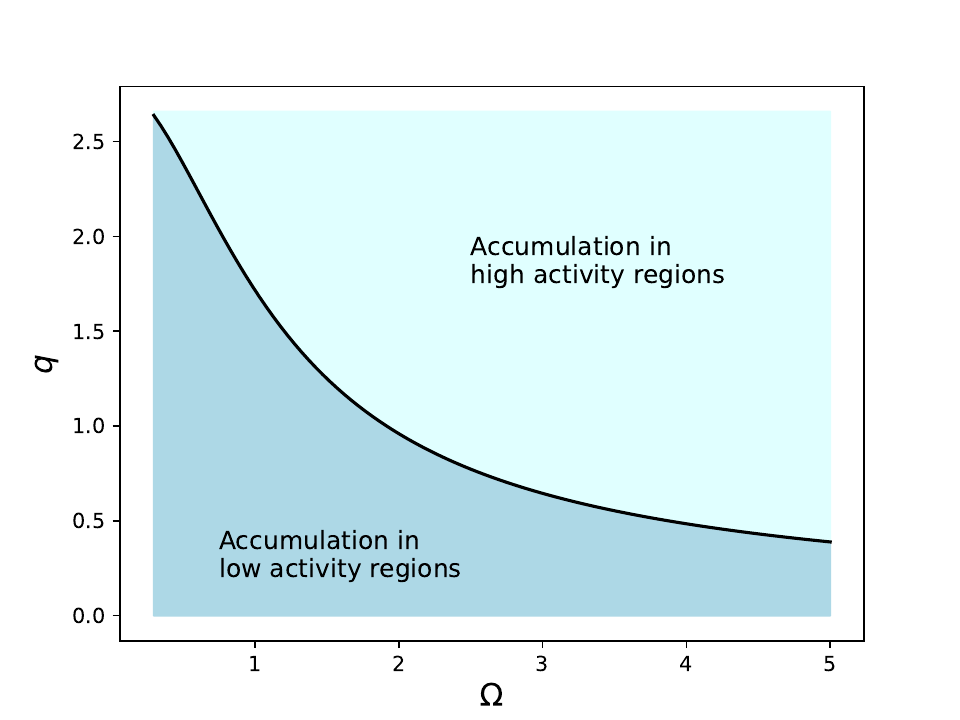}}
  \caption{Left: Comparison of steady state density distributions for the two extreme cases of the chiral torque $\Omega$. In particular, when the chiral active particle is spinning slowly (red), and the passive particle is twice the size of the chiral active particle, the composite accumulates in low-activity regions. Whereas, when the chiral active particle is spinning extremely fast (purple), the composite accumulates in high-activity regions even for a passive particle half the size of the chiral active particle. The symbols represent simulations with the same parameters and the sinusoidal activity field as in Fig.~\ref{fig:steady_state_density_spring_zero_restlenght}. They are in excellent agreement with the theoretical predictions (solid lines). Right: State diagram in $(q,\Omega)$ for $\gamma D_r/k = 0.75$. The solid black curve represents the minimum size of the passive particle needed to reverse the accumulation behavior of the active-passive composite. Particularly at large values of the chiral torque, a small passive particle can facilitate this transition.}
\label{fig:omega_extreme_cases_q_omega_state_diagram}
\end{figure*}
The quasi-stationarity of $\boldsymbol{\sigma}(\boldsymbol{R,r},t)$, combined with the assumption of small spatial variations in the activity field, allows for the closure of the hierarchy of the system of equations without requiring information from higher-order modes. These are known as the adiabatic approximation and the small gradients approximation, respectively. Furthermore, we integrate out the bond coordinate $\boldsymbol{r}$ from the continuity equation in Eq.~\eqref{eq:phi_dynamics_transformed} and derive an effective drift-diffusion equation for the probability density $\rho(\boldsymbol{R},t) = 2\pi \int \dd\boldsymbol{r}\, \phi(\boldsymbol{R},\boldsymbol{r},t)$ of the collective coordinate of the active-passive composite:
\begin{equation}
\begin{split}
    \frac{\partial \rho(\boldsymbol{R},t)}{\partial t} &=-\nabla_{\boldsymbol{R}}\cdot\boldsymbol{J},\\
    & = -\nabla_{\boldsymbol{R}}\cdot\bigg[\boldsymbol{V}(\boldsymbol{R})\rho(\boldsymbol{R}) - \nabla_{ \boldsymbol{R}}  \cdot \left(\boldsymbol{\mathds{D}}(\boldsymbol{R})\rho(\boldsymbol{R})\right)\bigg].
\end{split}    
\label{eq:drift_diffusiom_equation}
\end{equation}
%\newpage
Here, the effective diffusion coefficient $\boldsymbol{\mathds{D}}(\boldsymbol{R})$ now depends on $\boldsymbol{R}$ and is given by:
\begin{equation}
  \boldsymbol{\mathds{D}}(\boldsymbol{R}) = \dfrac{1}{1+q}\frac{T}{\gamma}\boldsymbol{\mathds{1}}+ \dfrac{1}{(1+q)^2}\dfrac{1}{2\gamma^2}f_{s}^{2}(\boldsymbol{R})\boldsymbol{\mathds{L}}^{T}.
\label{eq:effective_diffusion}  
\end{equation}  
Notably, $\boldsymbol{\mathds{D}}(\boldsymbol{R})$ has an antisymmetric part, which is a hallmark of odd-diffusive systems\cite{hargus_odd_2021,kalz_collisions_2022,kalz_oscillatory_2024,hargus_flux_2024,kalz_field_2024}.
The effective drift $\boldsymbol{V}(\boldsymbol{R})$ can be written in terms of $\boldsymbol{\mathds{D}}(\boldsymbol{R})$ as:

\begin{equation}
\begin{split}
    \boldsymbol{V}(\boldsymbol{R}) &= \left(\boldsymbol{\mathds{1}}-\frac{1}{2}  \boldsymbol{\mathds{L}}\big[(1-q)\boldsymbol{\mathds{1}}+(1+q)\boldsymbol{\mathds{B}}\big]\boldsymbol{\mathds{L}}^{-1}\right)\nabla_{\boldsymbol{R}}\cdot \boldsymbol{\mathds{D}}(\boldsymbol{R}).
    \label{eq:effective_drift}
\end{split}
\end{equation}
where the matrix $\boldsymbol{\mathds{B}}$ reads
\begin{equation}
   \boldsymbol{\mathds{B}} = \frac{qk}{\gamma} \boldsymbol{\mathds{L}}\left(q\boldsymbol{\mathds{1}}+\frac{(1+q)k}{\gamma}\boldsymbol{\mathds{L}}\right)^{-1}. 
\end{equation}
%\bigskip

We consider the activity field to vary only along the $x$-direction in the remainder of this study. Then, symmetry arguments dictate that the steady state density also varies only along the $x$ coordinate. The stationary probability flux along this direction is then given by:
\begin{equation}
\begin{split}
\label{eq:x_flux}
    J_x(x) =-\frac{\epsilon}{2} \rho \partial_x \mathds{D}_{xx} - \mathds{D}_{xx} \partial_x \rho, 
\end{split}
\end{equation}
where $\mathds{D}_{xx} $ denotes the $xx$ element of the effective diffusion coefficient given in Eq.~\eqref{eq:effective_diffusion}. Furthermore, we show below that the quantity $\epsilon$, which we call the \emph{tactic parameter}, determines the accumulation behavior of the composite. Upon imposing a zero-flux condition along this direction, we obtain the steady state density:
\begin{equation}
  \rho(x) \propto \bigg[1+\dfrac{D_{R}}{D_{R}^{2}+\omega^2}\dfrac{1}{1+q}\dfrac{1}{2\gamma T}f_{s}^{2}(x)\bigg]^{-\displaystyle \epsilon/2}.
\label{eq:steady_state_density}
\end{equation}

We find that the steady state density of our active-passive composite is determined by the sign of the exponent $\epsilon$. When $\epsilon<0\,(>0)$, $\rho(x)$ follows the same (opposite) trend as $f_s(x)$, i.e., the composite accumulates where the activity is high (low). As shown in the following, the tactic parameter depends on the interaction $\boldsymbol{F}$ between the chiral active particle and the passive particle, the ratio of mobilities $q$ and the chiral torque $\omega$. Specifically, we consider two cases for the force $\boldsymbol{F}$ in our analytical treatment: force due to a spring with zero rest length and, force due to an infinitely stiff spring.
%\newpage

We first consider the case of a harmonic spring with zero rest length. In this case, the tactic parameter is given by the expression:
%\newpage
\begin{equation}
\begin{split}
  \epsilon = 1 - q\dfrac{(1+\Omega^2)(1+\tau)}{\Omega^2 + (1+\tau)^2},
\end{split}  
  \label{eq:epsilon_spring}
\end{equation}  
where we have introduced non-dimensional parameters
\begin{equation}
\begin{split}
    \Omega &= \omega\tau_p,\\
    \tau &= \frac{(1+q)k\tau_p}{q\gamma},
\end{split}
\end{equation}
which express the chiral torque and the spring relaxation time in units of the persistence time $\tau_p = D_R^{-1}$ of the active particle due to rotational diffusion, respectively.

The steady state density given by Eqs.~\eqref{eq:steady_state_density} and~\eqref{eq:epsilon_spring} is shown in Fig.~\ref{fig:steady_state_density_spring_zero_restlenght}, for the case of a sinusoidally varying activity field. The accumulation of the active-passive composite can be reversed from low-activity regions to high-activity regions by increasing the chiral torque $\Omega$. In the limit of vanishing chiral torque, the tactic parameter (Eq.~\eqref{eq:epsilon_spring}) reduces to
\begin{equation}
    \lim_{\Omega\rightarrow 0}\epsilon = 1 - \dfrac{q}{1+\tau},
\label{eq:epsilon_spring_Omega_0}
\end{equation}
%\newpage
\noindent as previously reported in the study~\cite{vuijk_chemotaxis_2021}, which focused on an active Brownian particle coupled to a passive particle. In this case, the passive particle has to be at least as large as the active particle ($q>1$) for the composite to accumulate in high-activity regions. However, when the active particle experiences a chiral torque, even composites with a passive particle much smaller than the active particle ($q<1$) can accumulate in high-activity regions, provided the chiral torque is sufficiently large. This finding is depicted in the left panel of Fig.~\ref{fig:omega_extreme_cases_q_omega_state_diagram}. This indicates that the accumulation behavior of the active-passive composite can be tuned by applying a chiral torque to the active particle or by changing the size of the passive particle. We show this in the right panel of Fig.~\ref{fig:omega_extreme_cases_q_omega_state_diagram}, where, for every value of the chiral torque, there exists a critical size of the passive particle beyond which the tactic behavior of the active-passive composite is reversed. 

\begin{figure}[b]
\centering
\includegraphics[width=9.25cm, height=7.25cm]{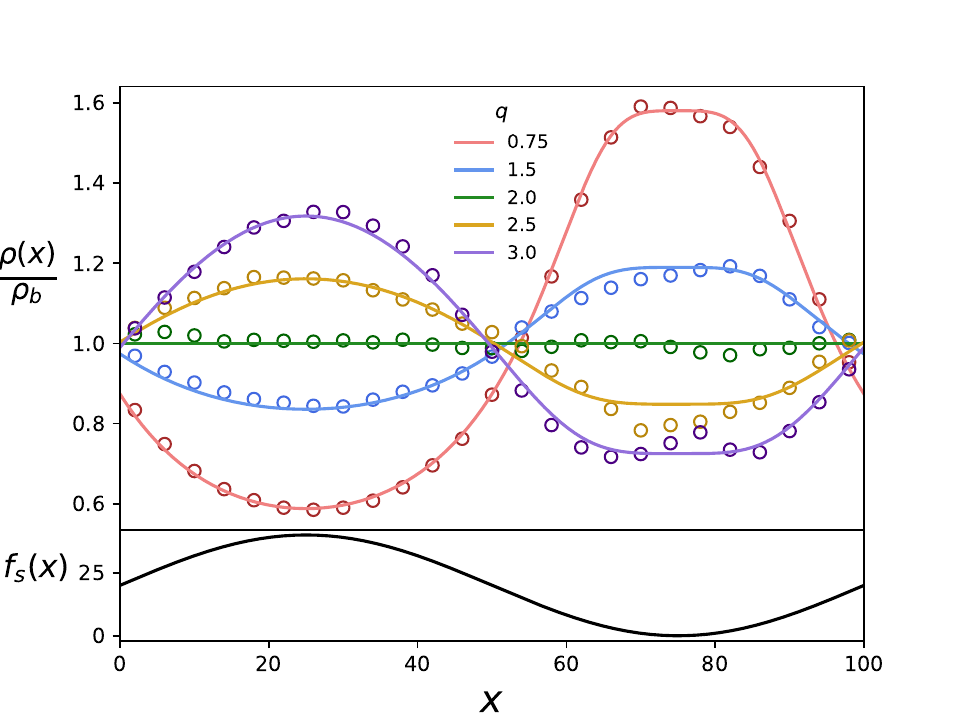}% Here is how to import EPS art
  \caption{Steady state density distribution (top panel) of the active-passive composites connected via an infinitely stiff spring for different sizes ($q$) of the passive particle. The solid lines are the analytical predictions (Eq.~\eqref{eq:steady_state_density} and~\eqref{eq:epsilon_stiff}) and the symbols are obtained from Langevin dynamics simulations of Eq.~\eqref{eq:chiral_active_passive_dimer_dynamics}. Unlike in Fig.~\ref{fig:steady_state_density_spring_zero_restlenght}, the size of the passive particle determines the steady state accumulation: the composite accumulates in regions of high-activity for larger passive particles and in regions of low-activity for smaller passive particles respectively. The bottom panel shows the activity field $f_s(x) = 20\big[1+\sin(2\pi x/L)\big]$ experienced by the chiral active particle. The $y$-axis in the top panel is normalized with the bulk density, defined as $\rho_b = 1/L$, where $L=100$ is the simulation box size with periodic boundary conditions. The parameters of the simulation are $k_{B}T=1.0$, $k=500.0$, $\gamma=1.0$, $D_{R} = 20.0$, $\Omega = 0.2$, $l_0=3.0$ and the integration time step $\Delta t=5D_R\times 10^{-6}$.}
\label{fig:steady_state_density_stiff_spring}  
\end{figure}

The preferential accumulation of the composite can be qualitatively understood as follows. Consider that the persistence time of the chiral active particle is smaller than the relaxation time of the spring connecting the two particles. The chiral active particle probes the neighborhood of the passive particle, locally sampling the activity gradients. This gives rise to a net pull on the passive particle towards the high-activity region. Since the persistence time of the active particle can be tuned via the chiral torque, i.e., it becomes smaller than the spring relaxation time with increasing $\Omega$, the tactic behavior of the composite can be switched from accumulation in low-activity regions to high-activity regions by increasing the chiral torque. It also follows that there is no tactic transition in the extreme case of $k \rightarrow \infty$ ($\epsilon \rightarrow 1$), in which case the spring relaxes instantaneously. This scenario corresponds to the chiral active particle and the passive particle being on top of each other.

Tuning the persistence time of the active particle via the chiral torque provides a handle on controlling the tactic behavior of composites. However, when fixing the distance between the chiral active particle and the passive particle via a rigid bond, the tactic behavior of the composite is independent of the persistence time and is determined by $q$ alone. The tactic parameter (See SM) then reads
\begin{equation}
  \epsilon = 1 - \frac{q}{2},
\label{eq:epsilon_stiff}
\end{equation}
same as reported in Ref. \cite{vuijk_chemotaxis_2021} for two dimensions. This is shown in Fig.~\ref{fig:steady_state_density_stiff_spring} for a sinusoidal variation of the activity field, where the transition in the accumulation behavior is obtained only by changing the ratio $q$.

\begin{figure}[b!]
    \centering
    \includegraphics[width=1\linewidth]{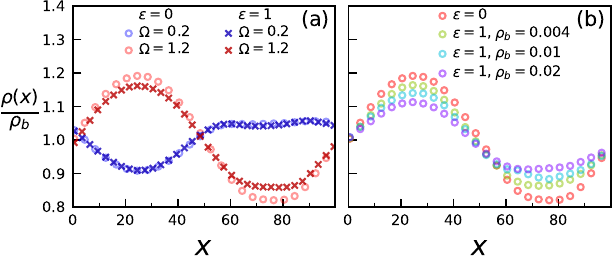}
    \caption{The steady state density distribution of active-passive composites with $q=2$ connected via a spring with $\ell_0=1.5$ and $k=30$. The $y$-axis is normalized by the bulk density $\rho_b$. The sinusoidal activity field is the same as in Fig.~\ref{fig:steady_state_density_spring_zero_restlenght}, given by $f_s(x) = 20 \left(1 + \sin\left({2\pi x}/{L}\right)\right)$. 
    Panel (a) compares the results for Case 1, where bonded particles interact via the WCA potential with $\varepsilon=1$. For reference, results for fully noninteracting dimers (i.e., neither bonded nor nonbonded interactions) with $\varepsilon=0$ are also shown. The accumulation behavior of the composite is influenced by the strength of the chiral torque $\Omega$. Panel (b) presents results for Case 2, where, in addition to bonded interactions, nonbonded dimers also interact via the WCA potential ($\varepsilon=1$). These results are compared with fully noninteracting dimers. The parameters of the simulation are $\Omega=1.2$, $k_{\rm B} T= 1.0$, $\gamma = 1.0$, $D_{R} = 10.0$, and the integration time step $\Delta t = 5D_R \times 10^{-6}$.}
    \label{fig:5}
\end{figure}

To determine whether chemotaxis persists in the presence of excluded volume effects, we performed simulations of dimers where bonded particles interact via the Weeks-Chandler-Anderson (WCA) potential, given by $U_{\rm WCA}(r) = 4\varepsilon \left[ \left(\frac{\sigma}{r}\right)^{12} - \left(\frac{\sigma}{r}\right)^6 \right] + \varepsilon$ for $r \leq 2^{1/6} \sigma$, and $U_{\rm WCA}(r) = 0$ for $r > 2^{1/6} \sigma$.
Here, \( \varepsilon \) represents the interaction strength, \( \sigma \) is the characteristic length scale of the interaction, and \( r \) denotes the center-to-center separation distance between two particles.
We considered two cases: (1) where nonbonded interactions are absent and (2) where nonbonded particles also interact via the WCA potential.

Figure~\ref{fig:5} presents the density profile along the $x$ axis for dimers with $q=2$, $k=30$, and $\ell_0=1.5$. Panel (a) corresponds to Case 1, where only bonded particles interact via the WCA potential. As expected, dimers with a chiral torque of $\Omega=0.2$ on the acitve particle accumulate in low-activity regions, and including WCA interactions between bonded particles has no observable effects. In contrast, for $\Omega=1.2$, where dimers accumulate in the high-activity regions, introducing WCA interactions between bonded particles slightly reduces the chemotactic response.
Panel (b) presents the results for Case 2.
In the dilute limit, chemotaxis remains robust, with dimers accumulating in high-activity regions. As the bulk density increases, particle collisions lead to reorientations, which consequently reduce the accumulation in the high-activity regions slightly.

\section{Conclusion \& Outlook} 
Directed motion toward regions of high concentration is a defining feature of living systems, spanning length scales from protein complexes to bacteria. In this work, we demonstrate that a minimal model system composed of a chiral active particle coupled to a passive particle, can exhibit preferential accumulation as an emergent property. Particularly, when the chiral active particle and the passive particle are connected by a rigid rod, the transition from accumulation in low-activity regions to high-activity regions depends solely on the size of the passive particle, similar to the previous studies~\cite{vuijk_chemotaxis_2021,muzzeddu_taxis_2023}. In contrast, when the coupling is mediated by a harmonic spring of zero rest length, the crossover is determined by the chiral torque. Additionally, we observe that this preferential accumulation is possible even when the chiral active particle is coupled to a small passive particle, provided the chiral torque is sufficiently large, which deviates from previous findings.
Our results demonstrate that even when excluded volume effects are considered, accumulation in high-activity regions remains robust in the dilute limit, further underscoring the effect of chirality as a control parameter. The mechanism underlying the preferential accumulation involves the anchoring effect of the passive particle, which allows the chiral active particle to explore the activity gradients and leads to the composite effectively drifting up the gradient. Moreover, our results show that when chiral active particles are connected into chains, they accumulate in high-activity regions at sufficiently high $\Omega$ for chain lengths exceeding three monomers (cf. Fig.~SI~1 of the SM). Previous theoretical studies suggested that directed motion can emerge when two chiral active particles with equal and opposite torques are coupled, creating a tug-of-war effect~\cite{muzzeddu_active_2022}. In contrast, we show that dimers with the same torque direction experience a cumulative chiral effect that suppresses chemotaxis, even at very high chiral torques. This can be attributed to the reduction of the persistence time of the active monomers in the presence of chirality, in which case short chains preferentially accumulate in low activity regions~\cite{muzzeddu_migration_2024}.

Chiral active matter is ubiquitous in nature, where chiral symmetry breaking at the cellular and subcellular levels contributes to bilateral asymmetry in organisms, organs, and tissues~\cite{hayashi2001, shibazaki2004, huang2012, taniguchi2011, savin2011}. At the cellular scale, active torques generated by the actomyosin cytoskeleton induce chiral symmetry breaking~\cite{naganathan2014}, while membrane-anchored motor proteins produce rotational forces that drive flagellar motion and cell propulsion~\cite{purcell1997, riedel_self-organized_2005}.
Beyond biological systems, chirality plays a crucial role in practical applications, enabling functionalities such as particle sorting~\cite{mijalkov2013, su2019, xu2022} and synchronization~\cite{levis2019, samatas2023}.
At the millimeter scale, vibrots provide a well-controlled experimental platform for studying chiral active matter~\cite{altshuler_vibrot_2013, scholz_ratcheting_2016, scholz_velocity_2017, broseghini_notched_2019, scholz_rotating_2018}, with tunable chiral torques achieved through bristle configurations. These systems can also be assembled into polymeric chains connected by springs~\cite{scholz_surfactants_2021}.
Chiral active granular particles have similarly been realized by breaking translational symmetry~\cite{dauchot2019, engbring2023}.
At the microscale, recent experiments have shown that amoeboid cell motility can be enhanced by coupling them to passive beads~\cite{lepro_optimal_2022}.
Our results offer a theoretical framework that can be tested in experimental systems where chirality emerges from symmetry breaking in body shape or self-propulsion mechanisms. Possible microscale realizations include active colloids, such as L-shaped particles confined using optical tweezers, or catalysis-coated synthetic swimmers~\cite{marchetti_hydrodynamics_2013, bechinger_active_2016, archer2015}.
Advances in fabrication techniques, such as glancing angle metal evaporation, now allow precise control over the spin rates of catalytic Janus swimmers, enabling systematic investigations of chiral active dynamics~\cite{archer2015}.

Our model does not account for the hydrodynamic interactions between the active-passive composite with the surrounding solvent, a class of models known as dry active matter models~\cite{marchetti_hydrodynamics_2013}. An intriguing extension of this work would involve incorporating hydrodynamic effects into the modeling, which we delegate to future studies.

\section{Acknowledgements}
A.S.~acknowledges support by the Deutsche Forschungsgemeinschaft (DFG) within Project No.~SH 1275/5-1. J.U.S.~thanks the cluster of excellence “Physics of Life” at TU Dresden for its support. B.V.~thanks S. Ravichandir for fruitful discussions, and A. Pandit for assistance in designing the schematic.

\bibliographystyle{apsrev4-1}
\bibliography{references.bib}

\begin{thebibliography}{78}%
\makeatletter
\providecommand \@ifxundefined [1]{%
 \@ifx{#1\undefined}
}%
\providecommand \@ifnum [1]{%
 \ifnum #1\expandafter \@firstoftwo
 \else \expandafter \@secondoftwo
 \fi
}%
\providecommand \@ifx [1]{%
 \ifx #1\expandafter \@firstoftwo
 \else \expandafter \@secondoftwo
 \fi
}%
\providecommand \natexlab [1]{#1}%
\providecommand \enquote  [1]{``#1''}%
\providecommand \bibnamefont  [1]{#1}%
\providecommand \bibfnamefont [1]{#1}%
\providecommand \citenamefont [1]{#1}%
\providecommand \href@noop [0]{\@secondoftwo}%
\providecommand \href [0]{\begingroup \@sanitize@url \@href}%
\providecommand \@href[1]{\@@startlink{#1}\@@href}%
\providecommand \@@href[1]{\endgroup#1\@@endlink}%
\providecommand \@sanitize@url [0]{\catcode `\\12\catcode `\$12\catcode
  `\&12\catcode `\#12\catcode `\^12\catcode `\_12\catcode `\%12\relax}%
\providecommand \@@startlink[1]{}%
\providecommand \@@endlink[0]{}%
\providecommand \url  [0]{\begingroup\@sanitize@url \@url }%
\providecommand \@url [1]{\endgroup\@href {#1}{\urlprefix }}%
\providecommand \urlprefix  [0]{URL }%
\providecommand \Eprint [0]{\href }%
\providecommand \doibase [0]{http://dx.doi.org/}%
\providecommand \selectlanguage [0]{\@gobble}%
\providecommand \bibinfo  [0]{\@secondoftwo}%
\providecommand \bibfield  [0]{\@secondoftwo}%
\providecommand \translation [1]{[#1]}%
\providecommand \BibitemOpen [0]{}%
\providecommand \bibitemStop [0]{}%
\providecommand \bibitemNoStop [0]{.\EOS\space}%
\providecommand \EOS [0]{\spacefactor3000\relax}%
\providecommand \BibitemShut  [1]{\csname bibitem#1\endcsname}%
\let\auto@bib@innerbib\@empty
%</preamble>
\bibitem [{\citenamefont {H{\"a}nggi}\ and\ \citenamefont
  {Marchesoni}(2009)}]{hanggi_artificial_2009}%
  \BibitemOpen
  \bibfield  {author} {\bibinfo {author} {\bibfnamefont {P.}~\bibnamefont
  {H{\"a}nggi}}\ and\ \bibinfo {author} {\bibfnamefont {F.}~\bibnamefont
  {Marchesoni}},\ }\href {\doibase 10.1103/RevModPhys.81.387} {\bibfield
  {journal} {\bibinfo  {journal} {Rev. Mod. Phys.}\ }\textbf {\bibinfo {volume}
  {81}},\ \bibinfo {pages} {387} (\bibinfo {year} {2009})}\BibitemShut
  {NoStop}%
\bibitem [{\citenamefont {Ramaswamy}(2010)}]{ramaswamy_mechanics_2010}%
  \BibitemOpen
  \bibfield  {author} {\bibinfo {author} {\bibfnamefont {S.}~\bibnamefont
  {Ramaswamy}},\ }\href {\doibase 10.1146/annurev-conmatphys-070909-104101}
  {\bibfield  {journal} {\bibinfo  {journal} {Annu. Rev. Condens. Matter
  Phys.}\ }\textbf {\bibinfo {volume} {1}},\ \bibinfo {pages} {323} (\bibinfo
  {year} {2010})}\BibitemShut {NoStop}%
\bibitem [{\citenamefont {Marchetti}\ \emph {et~al.}(2013)\citenamefont
  {Marchetti}, \citenamefont {Joanny}, \citenamefont {Ramaswamy}, \citenamefont
  {Liverpool}, \citenamefont {Prost}, \citenamefont {Rao},\ and\ \citenamefont
  {Simha}}]{marchetti_hydrodynamics_2013}%
  \BibitemOpen
  \bibfield  {author} {\bibinfo {author} {\bibfnamefont {M.~C.}\ \bibnamefont
  {Marchetti}}, \bibinfo {author} {\bibfnamefont {J.~F.}\ \bibnamefont
  {Joanny}}, \bibinfo {author} {\bibfnamefont {S.}~\bibnamefont {Ramaswamy}},
  \bibinfo {author} {\bibfnamefont {T.~B.}\ \bibnamefont {Liverpool}}, \bibinfo
  {author} {\bibfnamefont {J.}~\bibnamefont {Prost}}, \bibinfo {author}
  {\bibfnamefont {M.}~\bibnamefont {Rao}}, \ and\ \bibinfo {author}
  {\bibfnamefont {R.~A.}\ \bibnamefont {Simha}},\ }\href {\doibase
  10.1103/RevModPhys.85.1143} {\bibfield  {journal} {\bibinfo  {journal} {Rev.
  Mod. Phys.}\ }\textbf {\bibinfo {volume} {85}},\ \bibinfo {pages} {1143}
  (\bibinfo {year} {2013})}\BibitemShut {NoStop}%
\bibitem [{\citenamefont {Bechinger}\ \emph {et~al.}(2016)\citenamefont
  {Bechinger}, \citenamefont {Di~Leonardo}, \citenamefont {L{\"o}wen},
  \citenamefont {Reichhardt}, \citenamefont {Volpe},\ and\ \citenamefont
  {Volpe}}]{bechinger_active_2016}%
  \BibitemOpen
  \bibfield  {author} {\bibinfo {author} {\bibfnamefont {C.}~\bibnamefont
  {Bechinger}}, \bibinfo {author} {\bibfnamefont {R.}~\bibnamefont
  {Di~Leonardo}}, \bibinfo {author} {\bibfnamefont {H.}~\bibnamefont
  {L{\"o}wen}}, \bibinfo {author} {\bibfnamefont {C.}~\bibnamefont
  {Reichhardt}}, \bibinfo {author} {\bibfnamefont {G.}~\bibnamefont {Volpe}}, \
  and\ \bibinfo {author} {\bibfnamefont {G.}~\bibnamefont {Volpe}},\ }\href
  {\doibase 10.1103/RevModPhys.88.045006} {\bibfield  {journal} {\bibinfo
  {journal} {Rev. Mod. Phys.}\ }\textbf {\bibinfo {volume} {88}},\ \bibinfo
  {pages} {045006} (\bibinfo {year} {2016})}\BibitemShut {NoStop}%
\bibitem [{\citenamefont {Ramaswamy}(2017)}]{ramaswamy_active_2017}%
  \BibitemOpen
  \bibfield  {author} {\bibinfo {author} {\bibfnamefont {S.}~\bibnamefont
  {Ramaswamy}},\ }\href {\doibase 10.1088/1742-5468/aa6bc5} {\bibfield
  {journal} {\bibinfo  {journal} {J. Stat. Mech.}\ }\textbf {\bibinfo {volume}
  {2017}},\ \bibinfo {pages} {054002} (\bibinfo {year} {2017})}\BibitemShut
  {NoStop}%
\bibitem [{\citenamefont {Berg}(2004)}]{berg_e_2004}%
  \BibitemOpen
  \bibinfo {editor} {\bibfnamefont {H.~C.}\ \bibnamefont {Berg}},\ ed.,\ \href
  {\doibase 10.1007/b97370} {{\selectlanguage {en}\emph {\bibinfo {title} {E.
  coli in {Motion}}}}},\ Biological and {Medical} {Physics}, {Biomedical}
  {Engineering}\ (\bibinfo  {publisher} {Springer},\ \bibinfo {address} {New
  York, NY},\ \bibinfo {year} {2004})\BibitemShut {NoStop}%
\bibitem [{\citenamefont {Cates}(2012)}]{cates_diffusive_2012}%
  \BibitemOpen
  \bibfield  {author} {\bibinfo {author} {\bibfnamefont {M.~E.}\ \bibnamefont
  {Cates}},\ }\href {\doibase 10.1088/0034-4885/75/4/042601} {\bibfield
  {journal} {\bibinfo  {journal} {Rep. Prog. Phys.}\ }\textbf {\bibinfo
  {volume} {75}},\ \bibinfo {pages} {042601} (\bibinfo {year}
  {2012})}\BibitemShut {NoStop}%
\bibitem [{\citenamefont {Kromer}\ \emph {et~al.}(2020)\citenamefont {Kromer},
  \citenamefont {de~la Cruz},\ and\ \citenamefont
  {Friedrich}}]{kromer_chemokinetic_2020}%
  \BibitemOpen
  \bibfield  {author} {\bibinfo {author} {\bibfnamefont {J.~A.}\ \bibnamefont
  {Kromer}}, \bibinfo {author} {\bibfnamefont {N.}~\bibnamefont {de~la Cruz}},
  \ and\ \bibinfo {author} {\bibfnamefont {B.~M.}\ \bibnamefont {Friedrich}},\
  }\href {\doibase 10.1103/PhysRevLett.124.118101} {\bibfield  {journal}
  {\bibinfo  {journal} {Phys. Rev. Lett.}\ }\textbf {\bibinfo {volume} {124}},\
  \bibinfo {pages} {118101} (\bibinfo {year} {2020})}\BibitemShut {NoStop}%
\bibitem [{\citenamefont {Howse}\ \emph {et~al.}(2007)\citenamefont {Howse},
  \citenamefont {Jones}, \citenamefont {Ryan}, \citenamefont {Gough},
  \citenamefont {Vafabakhsh},\ and\ \citenamefont
  {Golestanian}}]{howse_self-motile_2007}%
  \BibitemOpen
  \bibfield  {author} {\bibinfo {author} {\bibfnamefont {J.~R.}\ \bibnamefont
  {Howse}}, \bibinfo {author} {\bibfnamefont {R.~A.~L.}\ \bibnamefont {Jones}},
  \bibinfo {author} {\bibfnamefont {A.~J.}\ \bibnamefont {Ryan}}, \bibinfo
  {author} {\bibfnamefont {T.}~\bibnamefont {Gough}}, \bibinfo {author}
  {\bibfnamefont {R.}~\bibnamefont {Vafabakhsh}}, \ and\ \bibinfo {author}
  {\bibfnamefont {R.}~\bibnamefont {Golestanian}},\ }\href {\doibase
  10.1103/PhysRevLett.99.048102} {\bibfield  {journal} {\bibinfo  {journal}
  {Phys. Rev. Lett.}\ }\textbf {\bibinfo {volume} {99}},\ \bibinfo {pages}
  {048102} (\bibinfo {year} {2007})}\BibitemShut {NoStop}%
\bibitem [{\citenamefont {Fraenkel}\ and\ \citenamefont
  {Gunn}(1941)}]{fraenkel_orientation_1941}%
  \BibitemOpen
  \bibfield  {author} {\bibinfo {author} {\bibfnamefont {G.~S.}\ \bibnamefont
  {Fraenkel}}\ and\ \bibinfo {author} {\bibfnamefont {D.~L.}\ \bibnamefont
  {Gunn}},\ }\href {\doibase 10.1093/aesa/34.3.690a} {\bibfield  {journal}
  {\bibinfo  {journal} {Ann. Entomol. Soc. Am.}\ }\textbf {\bibinfo {volume}
  {34}},\ \bibinfo {pages} {690} (\bibinfo {year} {1941})}\BibitemShut
  {NoStop}%
\bibitem [{\citenamefont {Schnitzer}(1993)}]{schnitzer_theory_1993}%
  \BibitemOpen
  \bibfield  {author} {\bibinfo {author} {\bibfnamefont {M.~J.}\ \bibnamefont
  {Schnitzer}},\ }\href {\doibase 10.1103/PhysRevE.48.2553} {\bibfield
  {journal} {\bibinfo  {journal} {Phys. Rev. E}\ }\textbf {\bibinfo {volume}
  {48}},\ \bibinfo {pages} {2553} (\bibinfo {year} {1993})}\BibitemShut
  {NoStop}%
\bibitem [{\citenamefont {Sharma}\ and\ \citenamefont
  {Brader}(2017)}]{sharma_brownian_2017}%
  \BibitemOpen
  \bibfield  {author} {\bibinfo {author} {\bibfnamefont {A.}~\bibnamefont
  {Sharma}}\ and\ \bibinfo {author} {\bibfnamefont {J.~M.}\ \bibnamefont
  {Brader}},\ }\href {\doibase 10.1103/PhysRevE.96.032604} {\bibfield
  {journal} {\bibinfo  {journal} {Phys. Rev. E}\ }\textbf {\bibinfo {volume}
  {96}},\ \bibinfo {pages} {032604} (\bibinfo {year} {2017})}\BibitemShut
  {NoStop}%
\bibitem [{\citenamefont {Caprini}\ \emph {et~al.}(2022)\citenamefont
  {Caprini}, \citenamefont {Marconi}, \citenamefont {Wittmann},\ and\
  \citenamefont {L{\"o}wen}}]{caprini_dynamics_2022}%
  \BibitemOpen
  \bibfield  {author} {\bibinfo {author} {\bibfnamefont {L.}~\bibnamefont
  {Caprini}}, \bibinfo {author} {\bibfnamefont {U.~M.~B.}\ \bibnamefont
  {Marconi}}, \bibinfo {author} {\bibfnamefont {R.}~\bibnamefont {Wittmann}}, \
  and\ \bibinfo {author} {\bibfnamefont {H.}~\bibnamefont {L{\"o}wen}},\ }\href
  {\doibase 10.1039/D1SM01648B} {\bibfield  {journal} {\bibinfo  {journal}
  {Soft Matter}\ }\textbf {\bibinfo {volume} {18}},\ \bibinfo {pages} {1412}
  (\bibinfo {year} {2022})}\BibitemShut {NoStop}%
\bibitem [{\citenamefont {Keller}\ and\ \citenamefont
  {Segel}(1971)}]{keller_model_1971}%
  \BibitemOpen
  \bibfield  {author} {\bibinfo {author} {\bibfnamefont {E.~F.}\ \bibnamefont
  {Keller}}\ and\ \bibinfo {author} {\bibfnamefont {L.~A.}\ \bibnamefont
  {Segel}},\ }\href {\doibase 10.1016/0022-5193(71)90050-6} {\bibfield
  {journal} {\bibinfo  {journal} {J. Theor. Biol.}\ }\textbf {\bibinfo {volume}
  {30}},\ \bibinfo {pages} {225} (\bibinfo {year} {1971})}\BibitemShut
  {NoStop}%
\bibitem [{\citenamefont {Liebchen}\ and\ \citenamefont
  {L{\"o}wen}(2018)}]{liebchen_synthetic_2018}%
  \BibitemOpen
  \bibfield  {author} {\bibinfo {author} {\bibfnamefont {B.}~\bibnamefont
  {Liebchen}}\ and\ \bibinfo {author} {\bibfnamefont {H.}~\bibnamefont
  {L{\"o}wen}},\ }\href {\doibase 10.1021/acs.accounts.8b00215} {\bibfield
  {journal} {\bibinfo  {journal} {Acc. Chem. Res.}\ }\textbf {\bibinfo {volume}
  {51}},\ \bibinfo {pages} {2982} (\bibinfo {year} {2018})}\BibitemShut
  {NoStop}%
\bibitem [{\citenamefont {Stark}(2018)}]{stark_artificial_2018}%
  \BibitemOpen
  \bibfield  {author} {\bibinfo {author} {\bibfnamefont {H.}~\bibnamefont
  {Stark}},\ }\href {\doibase 10.1021/acs.accounts.8b00259} {\bibfield
  {journal} {\bibinfo  {journal} {Acc. Chem. Res.}\ }\textbf {\bibinfo {volume}
  {51}},\ \bibinfo {pages} {2681} (\bibinfo {year} {2018})}\BibitemShut
  {NoStop}%
\bibitem [{\citenamefont {Qian}\ \emph {et~al.}(2013)\citenamefont {Qian},
  \citenamefont {Montiel}, \citenamefont {Bregulla}, \citenamefont {Cichos},\
  and\ \citenamefont {Yang}}]{qian_harnessing_2013}%
  \BibitemOpen
  \bibfield  {author} {\bibinfo {author} {\bibfnamefont {B.}~\bibnamefont
  {Qian}}, \bibinfo {author} {\bibfnamefont {D.}~\bibnamefont {Montiel}},
  \bibinfo {author} {\bibfnamefont {A.}~\bibnamefont {Bregulla}}, \bibinfo
  {author} {\bibfnamefont {F.}~\bibnamefont {Cichos}}, \ and\ \bibinfo {author}
  {\bibfnamefont {H.}~\bibnamefont {Yang}},\ }\href {\doibase
  10.1039/C2SC21263C} {\bibfield  {journal} {\bibinfo  {journal} {Chem. Sci.}\
  }\textbf {\bibinfo {volume} {4}},\ \bibinfo {pages} {1420} (\bibinfo {year}
  {2013})}\BibitemShut {NoStop}%
\bibitem [{\citenamefont {Mano}\ \emph {et~al.}(2017)\citenamefont {Mano},
  \citenamefont {Delfau}, \citenamefont {Iwasawa},\ and\ \citenamefont
  {Sano}}]{mano_optimal_2017}%
  \BibitemOpen
  \bibfield  {author} {\bibinfo {author} {\bibfnamefont {T.}~\bibnamefont
  {Mano}}, \bibinfo {author} {\bibfnamefont {J.-B.}\ \bibnamefont {Delfau}},
  \bibinfo {author} {\bibfnamefont {J.}~\bibnamefont {Iwasawa}}, \ and\
  \bibinfo {author} {\bibfnamefont {M.}~\bibnamefont {Sano}},\ }\href {\doibase
  10.1073/pnas.1616013114} {\bibfield  {journal} {\bibinfo  {journal} {Proc.
  Natl. Acad. Sci. U.S.A.}\ }\textbf {\bibinfo {volume} {114}},\ \bibinfo
  {pages} {E2580} (\bibinfo {year} {2017})}\BibitemShut {NoStop}%
\bibitem [{\citenamefont {Massana-Cid}\ \emph {et~al.}(2022)\citenamefont
  {Massana-Cid}, \citenamefont {Maggi}, \citenamefont {Frangipane},\ and\
  \citenamefont {Di~Leonardo}}]{massana-cid_rectification_2022}%
  \BibitemOpen
  \bibfield  {author} {\bibinfo {author} {\bibfnamefont {H.}~\bibnamefont
  {Massana-Cid}}, \bibinfo {author} {\bibfnamefont {C.}~\bibnamefont {Maggi}},
  \bibinfo {author} {\bibfnamefont {G.}~\bibnamefont {Frangipane}}, \ and\
  \bibinfo {author} {\bibfnamefont {R.}~\bibnamefont {Di~Leonardo}},\ }\href
  {\doibase 10.1038/s41467-022-30201-1} {\bibfield  {journal} {\bibinfo
  {journal} {Nat. Commun.}\ }\textbf {\bibinfo {volume} {13}},\ \bibinfo
  {pages} {2740} (\bibinfo {year} {2022})}\BibitemShut {NoStop}%
\bibitem [{\citenamefont {Lozano}\ \emph {et~al.}(2016)\citenamefont {Lozano},
  \citenamefont {ten Hagen}, \citenamefont {L{\"o}wen},\ and\ \citenamefont
  {Bechinger}}]{lozano_phototaxis_2016}%
  \BibitemOpen
  \bibfield  {author} {\bibinfo {author} {\bibfnamefont {C.}~\bibnamefont
  {Lozano}}, \bibinfo {author} {\bibfnamefont {B.}~\bibnamefont {ten Hagen}},
  \bibinfo {author} {\bibfnamefont {H.}~\bibnamefont {L{\"o}wen}}, \ and\
  \bibinfo {author} {\bibfnamefont {C.}~\bibnamefont {Bechinger}},\ }\href
  {\doibase 10.1038/ncomms12828} {\bibfield  {journal} {\bibinfo  {journal}
  {Nat. Commun.}\ }\textbf {\bibinfo {volume} {7}},\ \bibinfo {pages} {12828}
  (\bibinfo {year} {2016})}\BibitemShut {NoStop}%
\bibitem [{\citenamefont {Lozano}\ and\ \citenamefont
  {Bechinger}(2019)}]{lozano_diffusing_2019}%
  \BibitemOpen
  \bibfield  {author} {\bibinfo {author} {\bibfnamefont {C.}~\bibnamefont
  {Lozano}}\ and\ \bibinfo {author} {\bibfnamefont {C.}~\bibnamefont
  {Bechinger}},\ }\href {\doibase 10.1038/s41467-019-10535-z} {\bibfield
  {journal} {\bibinfo  {journal} {Nat Commun}\ }\textbf {\bibinfo {volume}
  {10}},\ \bibinfo {pages} {2495} (\bibinfo {year} {2019})}\BibitemShut
  {NoStop}%
\bibitem [{\citenamefont {Martinez-Pedrero}\ and\ \citenamefont
  {Tierno}(2015)}]{martinez-pedrero_magnetic_2015}%
  \BibitemOpen
  \bibfield  {author} {\bibinfo {author} {\bibfnamefont {F.}~\bibnamefont
  {Martinez-Pedrero}}\ and\ \bibinfo {author} {\bibfnamefont {P.}~\bibnamefont
  {Tierno}},\ }\href {\doibase 10.1103/PhysRevApplied.3.051003} {\bibfield
  {journal} {\bibinfo  {journal} {Phys. Rev. Appl.}\ }\textbf {\bibinfo
  {volume} {3}},\ \bibinfo {pages} {051003} (\bibinfo {year}
  {2015})}\BibitemShut {NoStop}%
\bibitem [{\citenamefont {Geiseler}\ \emph {et~al.}(2017)\citenamefont
  {Geiseler}, \citenamefont {Hänggi},\ and\ \citenamefont
  {Marchesoni}}]{geiseler_self-polarizing_2017}%
  \BibitemOpen
  \bibfield  {author} {\bibinfo {author} {\bibfnamefont {A.}~\bibnamefont
  {Geiseler}}, \bibinfo {author} {\bibfnamefont {P.}~\bibnamefont {Hänggi}}, \
  and\ \bibinfo {author} {\bibfnamefont {F.}~\bibnamefont {Marchesoni}},\
  }\href {\doibase 10.1038/srep41884} {\bibfield  {journal} {\bibinfo
  {journal} {Scientific Reports}\ }\textbf {\bibinfo {volume} {7}},\ \bibinfo
  {pages} {41884} (\bibinfo {year} {2017})}\BibitemShut {NoStop}%
\bibitem [{\citenamefont {Geiseler}\ \emph {et~al.}(2016)\citenamefont
  {Geiseler}, \citenamefont {Hänggi}, \citenamefont {Marchesoni},
  \citenamefont {Mulhern},\ and\ \citenamefont
  {Savel'ev}}]{geiseler_chemotaxis_2016}%
  \BibitemOpen
  \bibfield  {author} {\bibinfo {author} {\bibfnamefont {A.}~\bibnamefont
  {Geiseler}}, \bibinfo {author} {\bibfnamefont {P.}~\bibnamefont {Hänggi}},
  \bibinfo {author} {\bibfnamefont {F.}~\bibnamefont {Marchesoni}}, \bibinfo
  {author} {\bibfnamefont {C.}~\bibnamefont {Mulhern}}, \ and\ \bibinfo
  {author} {\bibfnamefont {S.}~\bibnamefont {Savel'ev}},\ }\href {\doibase
  10.1103/PhysRevE.94.012613} {\bibfield  {journal} {\bibinfo  {journal}
  {Physical Review E}\ }\textbf {\bibinfo {volume} {94}},\ \bibinfo {pages}
  {012613} (\bibinfo {year} {2016})}\BibitemShut {NoStop}%
\bibitem [{\citenamefont {Vuijk}\ \emph {et~al.}(2021)\citenamefont {Vuijk},
  \citenamefont {Merlitz}, \citenamefont {Lang}, \citenamefont {Sharma},\ and\
  \citenamefont {Sommer}}]{vuijk_chemotaxis_2021}%
  \BibitemOpen
  \bibfield  {author} {\bibinfo {author} {\bibfnamefont {H.~D.}\ \bibnamefont
  {Vuijk}}, \bibinfo {author} {\bibfnamefont {H.}~\bibnamefont {Merlitz}},
  \bibinfo {author} {\bibfnamefont {M.}~\bibnamefont {Lang}}, \bibinfo {author}
  {\bibfnamefont {A.}~\bibnamefont {Sharma}}, \ and\ \bibinfo {author}
  {\bibfnamefont {J.-U.}\ \bibnamefont {Sommer}},\ }\href {\doibase
  10.1103/PhysRevLett.126.208102} {\bibfield  {journal} {\bibinfo  {journal}
  {Phys. Rev. Lett.}\ }\textbf {\bibinfo {volume} {126}},\ \bibinfo {pages}
  {208102} (\bibinfo {year} {2021})}\BibitemShut {NoStop}%
\bibitem [{\citenamefont {Vuijk}\ \emph {et~al.}(2022)\citenamefont {Vuijk},
  \citenamefont {Klempahn}, \citenamefont {Merlitz}, \citenamefont {Sommer},\
  and\ \citenamefont {Sharma}}]{vuijk_active_2022}%
  \BibitemOpen
  \bibfield  {author} {\bibinfo {author} {\bibfnamefont {H.~D.}\ \bibnamefont
  {Vuijk}}, \bibinfo {author} {\bibfnamefont {S.}~\bibnamefont {Klempahn}},
  \bibinfo {author} {\bibfnamefont {H.}~\bibnamefont {Merlitz}}, \bibinfo
  {author} {\bibfnamefont {J.-U.}\ \bibnamefont {Sommer}}, \ and\ \bibinfo
  {author} {\bibfnamefont {A.}~\bibnamefont {Sharma}},\ }\href {\doibase
  10.1103/PhysRevE.106.014617} {\bibfield  {journal} {\bibinfo  {journal}
  {Phys. Rev. E}\ }\textbf {\bibinfo {volume} {106}},\ \bibinfo {pages}
  {014617} (\bibinfo {year} {2022})}\BibitemShut {NoStop}%
\bibitem [{\citenamefont {Muzzeddu}\ \emph {et~al.}(2024)\citenamefont
  {Muzzeddu}, \citenamefont {Gambassi}, \citenamefont {Sommer},\ and\
  \citenamefont {Sharma}}]{muzzeddu_migration_2024}%
  \BibitemOpen
  \bibfield  {author} {\bibinfo {author} {\bibfnamefont {P.~L.}\ \bibnamefont
  {Muzzeddu}}, \bibinfo {author} {\bibfnamefont {A.}~\bibnamefont {Gambassi}},
  \bibinfo {author} {\bibfnamefont {J.-U.}\ \bibnamefont {Sommer}}, \ and\
  \bibinfo {author} {\bibfnamefont {A.}~\bibnamefont {Sharma}},\ }\href
  {\doibase 10.1103/PhysRevLett.133.118102} {\bibfield  {journal} {\bibinfo
  {journal} {Phys. Rev. Lett.}\ }\textbf {\bibinfo {volume} {133}},\ \bibinfo
  {pages} {118102} (\bibinfo {year} {2024})}\BibitemShut {NoStop}%
\bibitem [{\citenamefont {Liebchen}\ and\ \citenamefont
  {Levis}(2022)}]{liebchen_chiral_2022}%
  \BibitemOpen
  \bibfield  {author} {\bibinfo {author} {\bibfnamefont {B.}~\bibnamefont
  {Liebchen}}\ and\ \bibinfo {author} {\bibfnamefont {D.}~\bibnamefont
  {Levis}},\ }\href {\doibase 10.1209/0295-5075/ac8f69} {\bibfield  {journal}
  {\bibinfo  {journal} {EPL}\ }\textbf {\bibinfo {volume} {139}},\ \bibinfo
  {pages} {67001} (\bibinfo {year} {2022})}\BibitemShut {NoStop}%
\bibitem [{\citenamefont {L{\"o}wen}(2016)}]{lowen_chirality_2016}%
  \BibitemOpen
  \bibfield  {author} {\bibinfo {author} {\bibfnamefont {H.}~\bibnamefont
  {L{\"o}wen}},\ }\href {\doibase 10.1140/epjst/e2016-60054-6} {\bibfield
  {journal} {\bibinfo  {journal} {Eur. Phys. J. Spec. Top.}\ }\textbf {\bibinfo
  {volume} {225}},\ \bibinfo {pages} {2319} (\bibinfo {year}
  {2016})}\BibitemShut {NoStop}%
\bibitem [{\citenamefont {CRENSHAW}(1996)}]{crenshaw_new_1996}%
  \BibitemOpen
  \bibfield  {author} {\bibinfo {author} {\bibfnamefont {H.~C.}\ \bibnamefont
  {CRENSHAW}},\ }\href {\doibase 10.1093/icb/36.6.608} {\bibfield  {journal}
  {\bibinfo  {journal} {Am. Zool.}\ }\textbf {\bibinfo {volume} {36}},\
  \bibinfo {pages} {608} (\bibinfo {year} {1996})}\BibitemShut {NoStop}%
\bibitem [{\citenamefont {Lauga}\ \emph {et~al.}(2006)\citenamefont {Lauga},
  \citenamefont {DiLuzio}, \citenamefont {Whitesides},\ and\ \citenamefont
  {Stone}}]{lauga_swimming_2006}%
  \BibitemOpen
  \bibfield  {author} {\bibinfo {author} {\bibfnamefont {E.}~\bibnamefont
  {Lauga}}, \bibinfo {author} {\bibfnamefont {W.~R.}\ \bibnamefont {DiLuzio}},
  \bibinfo {author} {\bibfnamefont {G.~M.}\ \bibnamefont {Whitesides}}, \ and\
  \bibinfo {author} {\bibfnamefont {H.~A.}\ \bibnamefont {Stone}},\ }\href
  {\doibase 10.1529/biophysj.105.069401} {\bibfield  {journal} {\bibinfo
  {journal} {Biophys. J.}\ }\textbf {\bibinfo {volume} {90}},\ \bibinfo {pages}
  {400} (\bibinfo {year} {2006})}\BibitemShut {NoStop}%
\bibitem [{\citenamefont {Riedel}\ \emph {et~al.}(2005)\citenamefont {Riedel},
  \citenamefont {Kruse},\ and\ \citenamefont
  {Howard}}]{riedel_self-organized_2005}%
  \BibitemOpen
  \bibfield  {author} {\bibinfo {author} {\bibfnamefont {I.~H.}\ \bibnamefont
  {Riedel}}, \bibinfo {author} {\bibfnamefont {K.}~\bibnamefont {Kruse}}, \
  and\ \bibinfo {author} {\bibfnamefont {J.}~\bibnamefont {Howard}},\ }\href
  {\doibase 10.1126/science.1110329} {\bibfield  {journal} {\bibinfo  {journal}
  {Science}\ }\textbf {\bibinfo {volume} {309}},\ \bibinfo {pages} {300}
  (\bibinfo {year} {2005})}\BibitemShut {NoStop}%
\bibitem [{\citenamefont {Friedrich}\ and\ \citenamefont
  {J{\"u}licher}(2007)}]{friedrich_chemotaxis_2007}%
  \BibitemOpen
  \bibfield  {author} {\bibinfo {author} {\bibfnamefont {B.~M.}\ \bibnamefont
  {Friedrich}}\ and\ \bibinfo {author} {\bibfnamefont {F.}~\bibnamefont
  {J{\"u}licher}},\ }\href {\doibase 10.1073/pnas.0703530104} {\bibfield
  {journal} {\bibinfo  {journal} {Proc. Natl. Acad. Sci. U.S.A.}\ }\textbf
  {\bibinfo {volume} {104}},\ \bibinfo {pages} {13256} (\bibinfo {year}
  {2007})}\BibitemShut {NoStop}%
\bibitem [{\citenamefont {Drescher}\ \emph {et~al.}(2009)\citenamefont
  {Drescher}, \citenamefont {Leptos}, \citenamefont {Tuval}, \citenamefont
  {Ishikawa}, \citenamefont {Pedley},\ and\ \citenamefont
  {Goldstein}}]{drescher_dancing_2009}%
  \BibitemOpen
  \bibfield  {author} {\bibinfo {author} {\bibfnamefont {K.}~\bibnamefont
  {Drescher}}, \bibinfo {author} {\bibfnamefont {K.~C.}\ \bibnamefont
  {Leptos}}, \bibinfo {author} {\bibfnamefont {I.}~\bibnamefont {Tuval}},
  \bibinfo {author} {\bibfnamefont {T.}~\bibnamefont {Ishikawa}}, \bibinfo
  {author} {\bibfnamefont {T.~J.}\ \bibnamefont {Pedley}}, \ and\ \bibinfo
  {author} {\bibfnamefont {R.~E.}\ \bibnamefont {Goldstein}},\ }\href {\doibase
  10.1103/PhysRevLett.102.168101} {\bibfield  {journal} {\bibinfo  {journal}
  {Phys. Rev. Lett.}\ }\textbf {\bibinfo {volume} {102}},\ \bibinfo {pages}
  {168101} (\bibinfo {year} {2009})}\BibitemShut {NoStop}%
\bibitem [{\citenamefont {K{\"u}mmel}\ \emph {et~al.}(2013)\citenamefont
  {K{\"u}mmel}, \citenamefont {ten Hagen}, \citenamefont {Wittkowski},
  \citenamefont {Buttinoni}, \citenamefont {Eichhorn}, \citenamefont {Volpe},
  \citenamefont {L{\"o}wen},\ and\ \citenamefont
  {Bechinger}}]{kummel_circular_2013}%
  \BibitemOpen
  \bibfield  {author} {\bibinfo {author} {\bibfnamefont {F.}~\bibnamefont
  {K{\"u}mmel}}, \bibinfo {author} {\bibfnamefont {B.}~\bibnamefont {ten
  Hagen}}, \bibinfo {author} {\bibfnamefont {R.}~\bibnamefont {Wittkowski}},
  \bibinfo {author} {\bibfnamefont {I.}~\bibnamefont {Buttinoni}}, \bibinfo
  {author} {\bibfnamefont {R.}~\bibnamefont {Eichhorn}}, \bibinfo {author}
  {\bibfnamefont {G.}~\bibnamefont {Volpe}}, \bibinfo {author} {\bibfnamefont
  {H.}~\bibnamefont {L{\"o}wen}}, \ and\ \bibinfo {author} {\bibfnamefont
  {C.}~\bibnamefont {Bechinger}},\ }\href {\doibase
  10.1103/PhysRevLett.110.198302} {\bibfield  {journal} {\bibinfo  {journal}
  {Phys. Rev. Lett.}\ }\textbf {\bibinfo {volume} {110}},\ \bibinfo {pages}
  {198302} (\bibinfo {year} {2013})}\BibitemShut {NoStop}%
\bibitem [{\citenamefont {Matsunaga}\ \emph {et~al.}(2019)\citenamefont
  {Matsunaga}, \citenamefont {Hamilton}, \citenamefont {Meng}, \citenamefont
  {Bukin}, \citenamefont {Martin}, \citenamefont {Ogrin}, \citenamefont
  {Yeomans},\ and\ \citenamefont {Golestanian}}]{matsunaga_controlling_2019}%
  \BibitemOpen
  \bibfield  {author} {\bibinfo {author} {\bibfnamefont {D.}~\bibnamefont
  {Matsunaga}}, \bibinfo {author} {\bibfnamefont {J.~K.}\ \bibnamefont
  {Hamilton}}, \bibinfo {author} {\bibfnamefont {F.}~\bibnamefont {Meng}},
  \bibinfo {author} {\bibfnamefont {N.}~\bibnamefont {Bukin}}, \bibinfo
  {author} {\bibfnamefont {E.~L.}\ \bibnamefont {Martin}}, \bibinfo {author}
  {\bibfnamefont {F.~Y.}\ \bibnamefont {Ogrin}}, \bibinfo {author}
  {\bibfnamefont {J.~M.}\ \bibnamefont {Yeomans}}, \ and\ \bibinfo {author}
  {\bibfnamefont {R.}~\bibnamefont {Golestanian}},\ }\href {\doibase
  10.1038/s41467-019-12665-w} {\bibfield  {journal} {\bibinfo  {journal} {Nat.
  Commun.}\ }\textbf {\bibinfo {volume} {10}},\ \bibinfo {pages} {4696}
  (\bibinfo {year} {2019})}\BibitemShut {NoStop}%
\bibitem [{\citenamefont {Arora}\ \emph {et~al.}(2021)\citenamefont {Arora},
  \citenamefont {Sood},\ and\ \citenamefont {Ganapathy}}]{arora_emergent_2021}%
  \BibitemOpen
  \bibfield  {author} {\bibinfo {author} {\bibfnamefont {P.}~\bibnamefont
  {Arora}}, \bibinfo {author} {\bibfnamefont {A.~K.}\ \bibnamefont {Sood}}, \
  and\ \bibinfo {author} {\bibfnamefont {R.}~\bibnamefont {Ganapathy}},\ }\href
  {\doibase 10.1126/sciadv.abd0331} {\bibfield  {journal} {\bibinfo  {journal}
  {Sci. Adv.}\ }\textbf {\bibinfo {volume} {7}},\ \bibinfo {pages} {eabd0331}
  (\bibinfo {year} {2021})}\BibitemShut {NoStop}%
\bibitem [{\citenamefont {Bililign}\ \emph {et~al.}(2022)\citenamefont
  {Bililign}, \citenamefont {Balboa~Usabiaga}, \citenamefont {Ganan},
  \citenamefont {Poncet}, \citenamefont {Soni}, \citenamefont {Magkiriadou},
  \citenamefont {Shelley}, \citenamefont {Bartolo},\ and\ \citenamefont
  {Irvine}}]{bililign_motile_2022}%
  \BibitemOpen
  \bibfield  {author} {\bibinfo {author} {\bibfnamefont {E.~S.}\ \bibnamefont
  {Bililign}}, \bibinfo {author} {\bibfnamefont {F.}~\bibnamefont
  {Balboa~Usabiaga}}, \bibinfo {author} {\bibfnamefont {Y.~A.}\ \bibnamefont
  {Ganan}}, \bibinfo {author} {\bibfnamefont {A.}~\bibnamefont {Poncet}},
  \bibinfo {author} {\bibfnamefont {V.}~\bibnamefont {Soni}}, \bibinfo {author}
  {\bibfnamefont {S.}~\bibnamefont {Magkiriadou}}, \bibinfo {author}
  {\bibfnamefont {M.~J.}\ \bibnamefont {Shelley}}, \bibinfo {author}
  {\bibfnamefont {D.}~\bibnamefont {Bartolo}}, \ and\ \bibinfo {author}
  {\bibfnamefont {W.~T.~M.}\ \bibnamefont {Irvine}},\ }\href {\doibase
  10.1038/s41567-021-01429-3} {\bibfield  {journal} {\bibinfo  {journal} {Nat.
  Phys.}\ }\textbf {\bibinfo {volume} {18}},\ \bibinfo {pages} {212} (\bibinfo
  {year} {2022})}\BibitemShut {NoStop}%
\bibitem [{\citenamefont {Caprini}\ and\ \citenamefont
  {Marconi}(2019)}]{caprini_active_2019}%
  \BibitemOpen
  \bibfield  {author} {\bibinfo {author} {\bibfnamefont {L.}~\bibnamefont
  {Caprini}}\ and\ \bibinfo {author} {\bibfnamefont {U.~M.~B.}\ \bibnamefont
  {Marconi}},\ }\href {\doibase 10.1039/C8SM02492H} {\bibfield  {journal}
  {\bibinfo  {journal} {Soft Matter}\ }\textbf {\bibinfo {volume} {15}},\
  \bibinfo {pages} {2627} (\bibinfo {year} {2019})}\BibitemShut {NoStop}%
\bibitem [{\citenamefont {Lei}\ \emph {et~al.}(2019)\citenamefont {Lei},
  \citenamefont {Ciamarra},\ and\ \citenamefont
  {Ni}}]{lei_nonequilibrium_2019}%
  \BibitemOpen
  \bibfield  {author} {\bibinfo {author} {\bibfnamefont {Q.-L.}\ \bibnamefont
  {Lei}}, \bibinfo {author} {\bibfnamefont {M.~P.}\ \bibnamefont {Ciamarra}}, \
  and\ \bibinfo {author} {\bibfnamefont {R.}~\bibnamefont {Ni}},\ }\href
  {\doibase 10.1126/sciadv.aau7423} {\bibfield  {journal} {\bibinfo  {journal}
  {Science Advances}\ }\textbf {\bibinfo {volume} {5}},\ \bibinfo {pages}
  {eaau7423} (\bibinfo {year} {2019})}\BibitemShut {NoStop}%
\bibitem [{\citenamefont {Marconi}\ and\ \citenamefont
  {Caprini}(2025)}]{marconi_spontaneous_2025}%
  \BibitemOpen
  \bibfield  {author} {\bibinfo {author} {\bibfnamefont {U.~M.~B.}\
  \bibnamefont {Marconi}}\ and\ \bibinfo {author} {\bibfnamefont
  {L.}~\bibnamefont {Caprini}},\ }\href {\doibase 10.1039/D4SM01426J}
  {\bibfield  {journal} {\bibinfo  {journal} {Soft Matter}\ } (\bibinfo {year}
  {2025}),\ 10.1039/D4SM01426J}\BibitemShut {NoStop}%
\bibitem [{\citenamefont {Ventejou}\ \emph {et~al.}(2021)\citenamefont
  {Ventejou}, \citenamefont {Chat{\'e}}, \citenamefont {Montagne},\ and\
  \citenamefont {Shi}}]{ventejou_susceptibility_2021}%
  \BibitemOpen
  \bibfield  {author} {\bibinfo {author} {\bibfnamefont {B.}~\bibnamefont
  {Ventejou}}, \bibinfo {author} {\bibfnamefont {H.}~\bibnamefont {Chat{\'e}}},
  \bibinfo {author} {\bibfnamefont {R.}~\bibnamefont {Montagne}}, \ and\
  \bibinfo {author} {\bibfnamefont {X.-q.}\ \bibnamefont {Shi}},\ }\href
  {\doibase 10.1103/PhysRevLett.127.238001} {\bibfield  {journal} {\bibinfo
  {journal} {Phys. Rev. Lett.}\ }\textbf {\bibinfo {volume} {127}},\ \bibinfo
  {pages} {238001} (\bibinfo {year} {2021})}\BibitemShut {NoStop}%
\bibitem [{\citenamefont {Souslov}\ \emph {et~al.}(2017)\citenamefont
  {Souslov}, \citenamefont {Van~Zuiden}, \citenamefont {Bartolo},\ and\
  \citenamefont {Vitelli}}]{souslov2017}%
  \BibitemOpen
  \bibfield  {author} {\bibinfo {author} {\bibfnamefont {A.}~\bibnamefont
  {Souslov}}, \bibinfo {author} {\bibfnamefont {B.~C.}\ \bibnamefont
  {Van~Zuiden}}, \bibinfo {author} {\bibfnamefont {D.}~\bibnamefont {Bartolo}},
  \ and\ \bibinfo {author} {\bibfnamefont {V.}~\bibnamefont {Vitelli}},\ }\href
  {\doibase https://doi.org/10.1038/nphys4193} {\bibfield  {journal} {\bibinfo
  {journal} {Nat. Phys.}\ }\textbf {\bibinfo {volume} {13}},\ \bibinfo {pages}
  {1091} (\bibinfo {year} {2017})}\BibitemShut {NoStop}%
\bibitem [{\citenamefont {Dunajova}\ \emph {et~al.}(2023)\citenamefont
  {Dunajova}, \citenamefont {Mateu}, \citenamefont {Radler}, \citenamefont
  {Lim}, \citenamefont {Brandis}, \citenamefont {Velicky}, \citenamefont
  {Danzl}, \citenamefont {Wong}, \citenamefont {Elgeti}, \citenamefont
  {Hannezo} \emph {et~al.}}]{dunajova2023}%
  \BibitemOpen
  \bibfield  {author} {\bibinfo {author} {\bibfnamefont {Z.}~\bibnamefont
  {Dunajova}}, \bibinfo {author} {\bibfnamefont {B.~P.}\ \bibnamefont {Mateu}},
  \bibinfo {author} {\bibfnamefont {P.}~\bibnamefont {Radler}}, \bibinfo
  {author} {\bibfnamefont {K.}~\bibnamefont {Lim}}, \bibinfo {author}
  {\bibfnamefont {D.}~\bibnamefont {Brandis}}, \bibinfo {author} {\bibfnamefont
  {P.}~\bibnamefont {Velicky}}, \bibinfo {author} {\bibfnamefont {J.~G.}\
  \bibnamefont {Danzl}}, \bibinfo {author} {\bibfnamefont {R.~W.}\ \bibnamefont
  {Wong}}, \bibinfo {author} {\bibfnamefont {J.}~\bibnamefont {Elgeti}},
  \bibinfo {author} {\bibfnamefont {E.}~\bibnamefont {Hannezo}},  \emph
  {et~al.},\ }\href {\doibase https://doi.org/10.1038/s41567-023-02218-w}
  {\bibfield  {journal} {\bibinfo  {journal} {Nat. Phys.}\ }\textbf {\bibinfo
  {volume} {19}},\ \bibinfo {pages} {1916} (\bibinfo {year}
  {2023})}\BibitemShut {NoStop}%
\bibitem [{\citenamefont {Shi}\ \emph {et~al.}(2020)\citenamefont {Shi},
  \citenamefont {Quint}, \citenamefont {Grason}, \citenamefont {Gopinathan},\
  and\ \citenamefont {Huang}}]{shi2020}%
  \BibitemOpen
  \bibfield  {author} {\bibinfo {author} {\bibfnamefont {H.}~\bibnamefont
  {Shi}}, \bibinfo {author} {\bibfnamefont {D.~A.}\ \bibnamefont {Quint}},
  \bibinfo {author} {\bibfnamefont {G.~M.}\ \bibnamefont {Grason}}, \bibinfo
  {author} {\bibfnamefont {A.}~\bibnamefont {Gopinathan}}, \ and\ \bibinfo
  {author} {\bibfnamefont {K.~C.}\ \bibnamefont {Huang}},\ }\href {\doibase
  https://doi.org/10.1038/s41467-020-14752-9} {\bibfield  {journal} {\bibinfo
  {journal} {Nat. Commun.}\ }\textbf {\bibinfo {volume} {11}},\ \bibinfo
  {pages} {1408} (\bibinfo {year} {2020})}\BibitemShut {NoStop}%
\bibitem [{\citenamefont {Risken}(1996)}]{risken_fokker-planck_1996}%
  \BibitemOpen
  \bibfield  {author} {\bibinfo {author} {\bibfnamefont {H.}~\bibnamefont
  {Risken}},\ }\href {\doibase 10.1007/978-3-642-61544-3} {{\selectlanguage
  {en}\emph {\bibinfo {title} {The {Fokker}-{Planck} {Equation}: {Methods} of
  {Solution} and {Applications}}}}},\ edited by\ \bibinfo {editor}
  {\bibfnamefont {H.}~\bibnamefont {Haken}},\ \bibinfo {series} {Springer
  {Series} in {Synergetics}}, Vol.~\bibinfo {volume} {18}\ (\bibinfo
  {publisher} {Springer},\ \bibinfo {address} {Berlin, Heidelberg},\ \bibinfo
  {year} {1996})\BibitemShut {NoStop}%
\bibitem [{\citenamefont {Muzzeddu}\ \emph {et~al.}(2023)\citenamefont
  {Muzzeddu}, \citenamefont {Rold{\'a}n}, \citenamefont {Gambassi},\ and\
  \citenamefont {Sharma}}]{muzzeddu_taxis_2023}%
  \BibitemOpen
  \bibfield  {author} {\bibinfo {author} {\bibfnamefont {P.~L.}\ \bibnamefont
  {Muzzeddu}}, \bibinfo {author} {\bibfnamefont {{\'E}.}~\bibnamefont
  {Rold{\'a}n}}, \bibinfo {author} {\bibfnamefont {A.}~\bibnamefont
  {Gambassi}}, \ and\ \bibinfo {author} {\bibfnamefont {A.}~\bibnamefont
  {Sharma}},\ }\href {\doibase 10.1209/0295-5075/acd8e9} {\bibfield  {journal}
  {\bibinfo  {journal} {EPL}\ }\textbf {\bibinfo {volume} {142}},\ \bibinfo
  {pages} {67001} (\bibinfo {year} {2023})}\BibitemShut {NoStop}%
\bibitem [{\citenamefont {Muzzeddu}\ \emph {et~al.}(2022)\citenamefont
  {Muzzeddu}, \citenamefont {Vuijk}, \citenamefont {L{\"o}wen}, \citenamefont
  {Sommer},\ and\ \citenamefont {Sharma}}]{muzzeddu_active_2022}%
  \BibitemOpen
  \bibfield  {author} {\bibinfo {author} {\bibfnamefont {P.~L.}\ \bibnamefont
  {Muzzeddu}}, \bibinfo {author} {\bibfnamefont {H.~D.}\ \bibnamefont {Vuijk}},
  \bibinfo {author} {\bibfnamefont {H.}~\bibnamefont {L{\"o}wen}}, \bibinfo
  {author} {\bibfnamefont {J.-U.}\ \bibnamefont {Sommer}}, \ and\ \bibinfo
  {author} {\bibfnamefont {A.}~\bibnamefont {Sharma}},\ }\href {\doibase
  10.1063/5.0109817} {\bibfield  {journal} {\bibinfo  {journal} {J. Chem.
  Phys.}\ }\textbf {\bibinfo {volume} {157}},\ \bibinfo {pages} {134902}
  (\bibinfo {year} {2022})}\BibitemShut {NoStop}%
\bibitem [{\citenamefont {Cates}\ and\ \citenamefont
  {Tailleur}(2013)}]{cates_when_2013}%
  \BibitemOpen
  \bibfield  {author} {\bibinfo {author} {\bibfnamefont {M.~E.}\ \bibnamefont
  {Cates}}\ and\ \bibinfo {author} {\bibfnamefont {J.}~\bibnamefont
  {Tailleur}},\ }\href {\doibase 10.1209/0295-5075/101/20010} {\bibfield
  {journal} {\bibinfo  {journal} {EPL}\ }\textbf {\bibinfo {volume} {101}},\
  \bibinfo {pages} {20010} (\bibinfo {year} {2013})}\BibitemShut {NoStop}%
\bibitem [{\citenamefont {Solon}\ \emph {et~al.}(2015)\citenamefont {Solon},
  \citenamefont {Cates},\ and\ \citenamefont {Tailleur}}]{solon_active_2015}%
  \BibitemOpen
  \bibfield  {author} {\bibinfo {author} {\bibfnamefont {A.~P.}\ \bibnamefont
  {Solon}}, \bibinfo {author} {\bibfnamefont {M.~E.}\ \bibnamefont {Cates}}, \
  and\ \bibinfo {author} {\bibfnamefont {J.}~\bibnamefont {Tailleur}},\ }\href
  {\doibase 10.1140/epjst/e2015-02457-0} {\bibfield  {journal} {\bibinfo
  {journal} {Eur. Phys. J. Spec. Top.}\ }\textbf {\bibinfo {volume} {224}},\
  \bibinfo {pages} {1231} (\bibinfo {year} {2015})}\BibitemShut {NoStop}%
\bibitem [{\citenamefont
  {Adeleke-Larodo}(2020)}]{adeleke-larodo_non-equilibrium_2020}%
  \BibitemOpen
  \bibfield  {author} {\bibinfo {author} {\bibfnamefont {T.}~\bibnamefont
  {Adeleke-Larodo}},\ }{\selectlanguage {English}\emph {\bibinfo {title}
  {Non-equilibrium dynamics of active enzymes}}},\ \href
  {https://ora.ox.ac.uk/objects/uuid:fcc43663-0ced-406b-8291-50999f912eff}
  {\bibinfo {type} {http://purl.org/dc/dcmitype/{Text}}},\ \bibinfo  {school}
  {University of Oxford} (\bibinfo {year} {2020})\BibitemShut {NoStop}%
\bibitem [{\citenamefont {Hargus}\ \emph {et~al.}(2021)\citenamefont {Hargus},
  \citenamefont {Epstein},\ and\ \citenamefont {Mandadapu}}]{hargus_odd_2021}%
  \BibitemOpen
  \bibfield  {author} {\bibinfo {author} {\bibfnamefont {C.}~\bibnamefont
  {Hargus}}, \bibinfo {author} {\bibfnamefont {J.~M.}\ \bibnamefont {Epstein}},
  \ and\ \bibinfo {author} {\bibfnamefont {K.~K.}\ \bibnamefont {Mandadapu}},\
  }\href {\doibase 10.1103/PhysRevLett.127.178001} {\bibfield  {journal}
  {\bibinfo  {journal} {Phys. Rev. Lett.}\ }\textbf {\bibinfo {volume} {127}},\
  \bibinfo {pages} {178001} (\bibinfo {year} {2021})}\BibitemShut {NoStop}%
\bibitem [{\citenamefont {Kalz}\ \emph {et~al.}(2022)\citenamefont {Kalz},
  \citenamefont {Vuijk}, \citenamefont {Abdoli}, \citenamefont {Sommer},
  \citenamefont {L{\"o}wen},\ and\ \citenamefont
  {Sharma}}]{kalz_collisions_2022}%
  \BibitemOpen
  \bibfield  {author} {\bibinfo {author} {\bibfnamefont {E.}~\bibnamefont
  {Kalz}}, \bibinfo {author} {\bibfnamefont {H.~D.}\ \bibnamefont {Vuijk}},
  \bibinfo {author} {\bibfnamefont {I.}~\bibnamefont {Abdoli}}, \bibinfo
  {author} {\bibfnamefont {J.-U.}\ \bibnamefont {Sommer}}, \bibinfo {author}
  {\bibfnamefont {H.}~\bibnamefont {L{\"o}wen}}, \ and\ \bibinfo {author}
  {\bibfnamefont {A.}~\bibnamefont {Sharma}},\ }\href {\doibase
  10.1103/PhysRevLett.129.090601} {\bibfield  {journal} {\bibinfo  {journal}
  {Phys. Rev. Lett.}\ }\textbf {\bibinfo {volume} {129}},\ \bibinfo {pages}
  {090601} (\bibinfo {year} {2022})}\BibitemShut {NoStop}%
\bibitem [{\citenamefont {Kalz}\ \emph
  {et~al.}(2024{\natexlab{a}})\citenamefont {Kalz}, \citenamefont {Vuijk},
  \citenamefont {Sommer}, \citenamefont {Metzler},\ and\ \citenamefont
  {Sharma}}]{kalz_oscillatory_2024}%
  \BibitemOpen
  \bibfield  {author} {\bibinfo {author} {\bibfnamefont {E.}~\bibnamefont
  {Kalz}}, \bibinfo {author} {\bibfnamefont {H.~D.}\ \bibnamefont {Vuijk}},
  \bibinfo {author} {\bibfnamefont {J.-U.}\ \bibnamefont {Sommer}}, \bibinfo
  {author} {\bibfnamefont {R.}~\bibnamefont {Metzler}}, \ and\ \bibinfo
  {author} {\bibfnamefont {A.}~\bibnamefont {Sharma}},\ }\href {\doibase
  10.1103/PhysRevLett.132.057102} {\bibfield  {journal} {\bibinfo  {journal}
  {Phys. Rev. Lett.}\ }\textbf {\bibinfo {volume} {132}},\ \bibinfo {pages}
  {057102} (\bibinfo {year} {2024}{\natexlab{a}})}\BibitemShut {NoStop}%
\bibitem [{\citenamefont {Hargus}\ \emph {et~al.}(2024)\citenamefont {Hargus},
  \citenamefont {Deshpande}, \citenamefont {Omar},\ and\ \citenamefont
  {Mandadapu}}]{hargus_flux_2024}%
  \BibitemOpen
  \bibfield  {author} {\bibinfo {author} {\bibfnamefont {C.}~\bibnamefont
  {Hargus}}, \bibinfo {author} {\bibfnamefont {A.}~\bibnamefont {Deshpande}},
  \bibinfo {author} {\bibfnamefont {A.~K.}\ \bibnamefont {Omar}}, \ and\
  \bibinfo {author} {\bibfnamefont {K.~K.}\ \bibnamefont {Mandadapu}},\ }\href
  {http://arxiv.org/abs/2405.08798} {{\selectlanguage {en}\enquote {\bibinfo
  {title} {The {Flux} {Hypothesis} for {Odd} {Transport} {Phenomena}},}\ }}
  (\bibinfo {year} {2024}),\ \bibinfo {note} {arXiv:2405.08798 [cond-mat,
  physics:physics]}\BibitemShut {NoStop}%
\bibitem [{\citenamefont {Kalz}\ \emph
  {et~al.}(2024{\natexlab{b}})\citenamefont {Kalz}, \citenamefont {Sharma},\
  and\ \citenamefont {Metzler}}]{kalz_field_2024}%
  \BibitemOpen
  \bibfield  {author} {\bibinfo {author} {\bibfnamefont {E.}~\bibnamefont
  {Kalz}}, \bibinfo {author} {\bibfnamefont {A.}~\bibnamefont {Sharma}}, \ and\
  \bibinfo {author} {\bibfnamefont {R.}~\bibnamefont {Metzler}},\ }\href
  {\doibase 10.1088/1751-8121/ad5089} {\bibfield  {journal} {\bibinfo
  {journal} {J. Phys. A: Math. Theor.}\ }\textbf {\bibinfo {volume} {57}},\
  \bibinfo {pages} {265002} (\bibinfo {year} {2024}{\natexlab{b}})}\BibitemShut
  {NoStop}%
\bibitem [{\citenamefont {Hayashi}\ and\ \citenamefont
  {Murakami}(2001)}]{hayashi2001}%
  \BibitemOpen
  \bibfield  {author} {\bibinfo {author} {\bibfnamefont {T.}~\bibnamefont
  {Hayashi}}\ and\ \bibinfo {author} {\bibfnamefont {R.}~\bibnamefont
  {Murakami}},\ }\href {\doibase
  https://doi.org/10.1046/j.1440-169x.2001.00574.x} {\bibfield  {journal}
  {\bibinfo  {journal} {Dev. Growth Differ.}\ }\textbf {\bibinfo {volume}
  {43}},\ \bibinfo {pages} {239} (\bibinfo {year} {2001})}\BibitemShut
  {NoStop}%
\bibitem [{\citenamefont {Shibazaki}\ \emph {et~al.}(2004)\citenamefont
  {Shibazaki}, \citenamefont {Shimizu},\ and\ \citenamefont
  {Kuroda}}]{shibazaki2004}%
  \BibitemOpen
  \bibfield  {author} {\bibinfo {author} {\bibfnamefont {Y.}~\bibnamefont
  {Shibazaki}}, \bibinfo {author} {\bibfnamefont {M.}~\bibnamefont {Shimizu}},
  \ and\ \bibinfo {author} {\bibfnamefont {R.}~\bibnamefont {Kuroda}},\ }\href
  {\doibase 10.1016/j.cub.2004.08.018} {\bibfield  {journal} {\bibinfo
  {journal} {Curr. Biol.}\ }\textbf {\bibinfo {volume} {14}},\ \bibinfo {pages}
  {1462} (\bibinfo {year} {2004})}\BibitemShut {NoStop}%
\bibitem [{\citenamefont {Huang}\ \emph {et~al.}(2012)\citenamefont {Huang},
  \citenamefont {Ehrhardt},\ and\ \citenamefont {Shaevitz}}]{huang2012}%
  \BibitemOpen
  \bibfield  {author} {\bibinfo {author} {\bibfnamefont {K.~C.}\ \bibnamefont
  {Huang}}, \bibinfo {author} {\bibfnamefont {D.~W.}\ \bibnamefont {Ehrhardt}},
  \ and\ \bibinfo {author} {\bibfnamefont {J.~W.}\ \bibnamefont {Shaevitz}},\
  }\href {\doibase https://doi.org/10.1016/j.mib.2012.11.002} {\bibfield
  {journal} {\bibinfo  {journal} {Curr. Opin. Microbiol.}\ }\textbf {\bibinfo
  {volume} {15}},\ \bibinfo {pages} {707} (\bibinfo {year} {2012})}\BibitemShut
  {NoStop}%
\bibitem [{\citenamefont {Taniguchi}\ \emph {et~al.}(2011)\citenamefont
  {Taniguchi}, \citenamefont {Maeda}, \citenamefont {Ando}, \citenamefont
  {Okumura}, \citenamefont {Nakazawa}, \citenamefont {Hatori}, \citenamefont
  {Nakamura}, \citenamefont {Hozumi}, \citenamefont {Fujiwara},\ and\
  \citenamefont {Matsuno}}]{taniguchi2011}%
  \BibitemOpen
  \bibfield  {author} {\bibinfo {author} {\bibfnamefont {K.}~\bibnamefont
  {Taniguchi}}, \bibinfo {author} {\bibfnamefont {R.}~\bibnamefont {Maeda}},
  \bibinfo {author} {\bibfnamefont {T.}~\bibnamefont {Ando}}, \bibinfo {author}
  {\bibfnamefont {T.}~\bibnamefont {Okumura}}, \bibinfo {author} {\bibfnamefont
  {N.}~\bibnamefont {Nakazawa}}, \bibinfo {author} {\bibfnamefont
  {R.}~\bibnamefont {Hatori}}, \bibinfo {author} {\bibfnamefont
  {M.}~\bibnamefont {Nakamura}}, \bibinfo {author} {\bibfnamefont
  {S.}~\bibnamefont {Hozumi}}, \bibinfo {author} {\bibfnamefont
  {H.}~\bibnamefont {Fujiwara}}, \ and\ \bibinfo {author} {\bibfnamefont
  {K.}~\bibnamefont {Matsuno}},\ }\href {\doibase 10.1126/science.1200940}
  {\bibfield  {journal} {\bibinfo  {journal} {Science}\ }\textbf {\bibinfo
  {volume} {333}},\ \bibinfo {pages} {339} (\bibinfo {year}
  {2011})}\BibitemShut {NoStop}%
\bibitem [{\citenamefont {Savin}\ \emph {et~al.}(2011)\citenamefont {Savin},
  \citenamefont {Kurpios}, \citenamefont {Shyer}, \citenamefont {Florescu},
  \citenamefont {Liang}, \citenamefont {Mahadevan},\ and\ \citenamefont
  {Tabin}}]{savin2011}%
  \BibitemOpen
  \bibfield  {author} {\bibinfo {author} {\bibfnamefont {T.}~\bibnamefont
  {Savin}}, \bibinfo {author} {\bibfnamefont {N.~A.}\ \bibnamefont {Kurpios}},
  \bibinfo {author} {\bibfnamefont {A.~E.}\ \bibnamefont {Shyer}}, \bibinfo
  {author} {\bibfnamefont {P.}~\bibnamefont {Florescu}}, \bibinfo {author}
  {\bibfnamefont {H.}~\bibnamefont {Liang}}, \bibinfo {author} {\bibfnamefont
  {L.}~\bibnamefont {Mahadevan}}, \ and\ \bibinfo {author} {\bibfnamefont
  {C.~J.}\ \bibnamefont {Tabin}},\ }\href {\doibase
  https://doi.org/10.1038/nature10277} {\bibfield  {journal} {\bibinfo
  {journal} {Nature}\ }\textbf {\bibinfo {volume} {476}},\ \bibinfo {pages}
  {57} (\bibinfo {year} {2011})}\BibitemShut {NoStop}%
\bibitem [{\citenamefont {Naganathan}\ \emph {et~al.}(2014)\citenamefont
  {Naganathan}, \citenamefont {F{\"u}rthauer}, \citenamefont {Nishikawa},
  \citenamefont {J{\"u}licher},\ and\ \citenamefont {Grill}}]{naganathan2014}%
  \BibitemOpen
  \bibfield  {author} {\bibinfo {author} {\bibfnamefont {S.~R.}\ \bibnamefont
  {Naganathan}}, \bibinfo {author} {\bibfnamefont {S.}~\bibnamefont
  {F{\"u}rthauer}}, \bibinfo {author} {\bibfnamefont {M.}~\bibnamefont
  {Nishikawa}}, \bibinfo {author} {\bibfnamefont {F.}~\bibnamefont
  {J{\"u}licher}}, \ and\ \bibinfo {author} {\bibfnamefont {S.~W.}\
  \bibnamefont {Grill}},\ }\href {\doibase https://doi.org/10.7554/eLife.04165}
  {\bibfield  {journal} {\bibinfo  {journal} {elife}\ }\textbf {\bibinfo
  {volume} {3}},\ \bibinfo {pages} {e04165} (\bibinfo {year}
  {2014})}\BibitemShut {NoStop}%
\bibitem [{\citenamefont {Purcell}(1997)}]{purcell1997}%
  \BibitemOpen
  \bibfield  {author} {\bibinfo {author} {\bibfnamefont {E.~M.}\ \bibnamefont
  {Purcell}},\ }\href {\doibase https://doi.org/10.1073/pnas.94.21.11307}
  {\bibfield  {journal} {\bibinfo  {journal} {Proc. Natl. Acad. Sci. U.S.A.}\
  }\textbf {\bibinfo {volume} {94}},\ \bibinfo {pages} {11307} (\bibinfo {year}
  {1997})}\BibitemShut {NoStop}%
\bibitem [{\citenamefont {Mijalkov}\ and\ \citenamefont
  {Volpe}(2013)}]{mijalkov2013}%
  \BibitemOpen
  \bibfield  {author} {\bibinfo {author} {\bibfnamefont {M.}~\bibnamefont
  {Mijalkov}}\ and\ \bibinfo {author} {\bibfnamefont {G.}~\bibnamefont
  {Volpe}},\ }\href {\doibase https://doi.org/10.1039/C3SM27923E} {\bibfield
  {journal} {\bibinfo  {journal} {Soft Matter}\ }\textbf {\bibinfo {volume}
  {9}},\ \bibinfo {pages} {6376} (\bibinfo {year} {2013})}\BibitemShut
  {NoStop}%
\bibitem [{\citenamefont {Su}\ \emph {et~al.}(2019)\citenamefont {Su},
  \citenamefont {Jiang},\ and\ \citenamefont {Hou}}]{su2019}%
  \BibitemOpen
  \bibfield  {author} {\bibinfo {author} {\bibfnamefont {J.}~\bibnamefont
  {Su}}, \bibinfo {author} {\bibfnamefont {H.}~\bibnamefont {Jiang}}, \ and\
  \bibinfo {author} {\bibfnamefont {Z.}~\bibnamefont {Hou}},\ }\href {\doibase
  https://doi.org/10.1039/C9SM01090D} {\bibfield  {journal} {\bibinfo
  {journal} {Soft Matter}\ }\textbf {\bibinfo {volume} {15}},\ \bibinfo {pages}
  {6830} (\bibinfo {year} {2019})}\BibitemShut {NoStop}%
\bibitem [{\citenamefont {Xu}\ \emph {et~al.}(2022)\citenamefont {Xu},
  \citenamefont {Li},\ and\ \citenamefont {Ai}}]{xu2022}%
  \BibitemOpen
  \bibfield  {author} {\bibinfo {author} {\bibfnamefont {G.-h.}\ \bibnamefont
  {Xu}}, \bibinfo {author} {\bibfnamefont {T.-C.}\ \bibnamefont {Li}}, \ and\
  \bibinfo {author} {\bibfnamefont {B.-q.}\ \bibnamefont {Ai}},\ }\href
  {\doibase https://doi.org/10.1016/j.physa.2022.128247} {\bibfield  {journal}
  {\bibinfo  {journal} {Physica A Stat. Mech. Appl.}\ }\textbf {\bibinfo
  {volume} {608}},\ \bibinfo {pages} {128247} (\bibinfo {year}
  {2022})}\BibitemShut {NoStop}%
\bibitem [{\citenamefont {Levis}\ \emph {et~al.}(2019)\citenamefont {Levis},
  \citenamefont {Pagonabarraga},\ and\ \citenamefont {Liebchen}}]{levis2019}%
  \BibitemOpen
  \bibfield  {author} {\bibinfo {author} {\bibfnamefont {D.}~\bibnamefont
  {Levis}}, \bibinfo {author} {\bibfnamefont {I.}~\bibnamefont
  {Pagonabarraga}}, \ and\ \bibinfo {author} {\bibfnamefont {B.}~\bibnamefont
  {Liebchen}},\ }\href {\doibase
  https://doi.org/10.1103/PhysRevResearch.1.023026} {\bibfield  {journal}
  {\bibinfo  {journal} {Phys. Rev. Res.}\ }\textbf {\bibinfo {volume} {1}},\
  \bibinfo {pages} {023026} (\bibinfo {year} {2019})}\BibitemShut {NoStop}%
\bibitem [{\citenamefont {Samatas}\ and\ \citenamefont
  {Lintuvuori}(2023)}]{samatas2023}%
  \BibitemOpen
  \bibfield  {author} {\bibinfo {author} {\bibfnamefont {S.}~\bibnamefont
  {Samatas}}\ and\ \bibinfo {author} {\bibfnamefont {J.}~\bibnamefont
  {Lintuvuori}},\ }\href {\doibase
  https://doi.org/10.1103/PhysRevLett.130.024001} {\bibfield  {journal}
  {\bibinfo  {journal} {Phys. Rev. Lett.}\ }\textbf {\bibinfo {volume} {130}},\
  \bibinfo {pages} {024001} (\bibinfo {year} {2023})}\BibitemShut {NoStop}%
\bibitem [{\citenamefont {Altshuler}\ \emph {et~al.}(2013)\citenamefont
  {Altshuler}, \citenamefont {Pastor}, \citenamefont {Garcimart{\'i}n},
  \citenamefont {Zuriguel},\ and\ \citenamefont
  {Maza}}]{altshuler_vibrot_2013}%
  \BibitemOpen
  \bibfield  {author} {\bibinfo {author} {\bibfnamefont {E.}~\bibnamefont
  {Altshuler}}, \bibinfo {author} {\bibfnamefont {J.~M.}\ \bibnamefont
  {Pastor}}, \bibinfo {author} {\bibfnamefont {A.}~\bibnamefont
  {Garcimart{\'i}n}}, \bibinfo {author} {\bibfnamefont {I.}~\bibnamefont
  {Zuriguel}}, \ and\ \bibinfo {author} {\bibfnamefont {D.}~\bibnamefont
  {Maza}},\ }\href {\doibase 10.1371/journal.pone.0067838} {\bibfield
  {journal} {\bibinfo  {journal} {PLOS ONE}\ }\textbf {\bibinfo {volume} {8}},\
  \bibinfo {pages} {e67838} (\bibinfo {year} {2013})}\BibitemShut {NoStop}%
\bibitem [{\citenamefont {Scholz}\ \emph {et~al.}(2016)\citenamefont {Scholz},
  \citenamefont {D{\textquoteright}Silva},\ and\ \citenamefont
  {P{\"o}schel}}]{scholz_ratcheting_2016}%
  \BibitemOpen
  \bibfield  {author} {\bibinfo {author} {\bibfnamefont {C.}~\bibnamefont
  {Scholz}}, \bibinfo {author} {\bibfnamefont {S.}~\bibnamefont
  {D{\textquoteright}Silva}}, \ and\ \bibinfo {author} {\bibfnamefont
  {T.}~\bibnamefont {P{\"o}schel}},\ }\href {\doibase
  10.1088/1367-2630/18/12/123001} {\bibfield  {journal} {\bibinfo  {journal}
  {New J. Phys.}\ }\textbf {\bibinfo {volume} {18}},\ \bibinfo {pages} {123001}
  (\bibinfo {year} {2016})}\BibitemShut {NoStop}%
\bibitem [{\citenamefont {Scholz}\ and\ \citenamefont
  {P{\"o}schel}(2017)}]{scholz_velocity_2017}%
  \BibitemOpen
  \bibfield  {author} {\bibinfo {author} {\bibfnamefont {C.}~\bibnamefont
  {Scholz}}\ and\ \bibinfo {author} {\bibfnamefont {T.}~\bibnamefont
  {P{\"o}schel}},\ }\href {\doibase 10.1103/PhysRevLett.118.198003} {\bibfield
  {journal} {\bibinfo  {journal} {Phys. Rev. Lett.}\ }\textbf {\bibinfo
  {volume} {118}},\ \bibinfo {pages} {198003} (\bibinfo {year}
  {2017})}\BibitemShut {NoStop}%
\bibitem [{\citenamefont {Broseghini}\ \emph {et~al.}(2019)\citenamefont
  {Broseghini}, \citenamefont {Ceccolini}, \citenamefont {Volpe},\ and\
  \citenamefont {Siboni}}]{broseghini_notched_2019}%
  \BibitemOpen
  \bibfield  {author} {\bibinfo {author} {\bibfnamefont {M.}~\bibnamefont
  {Broseghini}}, \bibinfo {author} {\bibfnamefont {C.}~\bibnamefont
  {Ceccolini}}, \bibinfo {author} {\bibfnamefont {C.~D.}\ \bibnamefont
  {Volpe}}, \ and\ \bibinfo {author} {\bibfnamefont {S.}~\bibnamefont
  {Siboni}},\ }\href {\doibase 10.1371/journal.pone.0218666} {\bibfield
  {journal} {\bibinfo  {journal} {PLOS ONE}\ }\textbf {\bibinfo {volume}
  {14}},\ \bibinfo {pages} {e0218666} (\bibinfo {year} {2019})}\BibitemShut
  {NoStop}%
\bibitem [{\citenamefont {Scholz}\ \emph {et~al.}(2018)\citenamefont {Scholz},
  \citenamefont {Engel},\ and\ \citenamefont
  {P{\"o}schel}}]{scholz_rotating_2018}%
  \BibitemOpen
  \bibfield  {author} {\bibinfo {author} {\bibfnamefont {C.}~\bibnamefont
  {Scholz}}, \bibinfo {author} {\bibfnamefont {M.}~\bibnamefont {Engel}}, \
  and\ \bibinfo {author} {\bibfnamefont {T.}~\bibnamefont {P{\"o}schel}},\
  }\href {\doibase 10.1038/s41467-018-03154-7} {\bibfield  {journal} {\bibinfo
  {journal} {Nat. Commun.}\ }\textbf {\bibinfo {volume} {9}},\ \bibinfo {pages}
  {931} (\bibinfo {year} {2018})}\BibitemShut {NoStop}%
\bibitem [{\citenamefont {Scholz}\ \emph {et~al.}(2021)\citenamefont {Scholz},
  \citenamefont {Ldov}, \citenamefont {P{\"o}schel}, \citenamefont {Engel},\
  and\ \citenamefont {L{\"o}wen}}]{scholz_surfactants_2021}%
  \BibitemOpen
  \bibfield  {author} {\bibinfo {author} {\bibfnamefont {C.}~\bibnamefont
  {Scholz}}, \bibinfo {author} {\bibfnamefont {A.}~\bibnamefont {Ldov}},
  \bibinfo {author} {\bibfnamefont {T.}~\bibnamefont {P{\"o}schel}}, \bibinfo
  {author} {\bibfnamefont {M.}~\bibnamefont {Engel}}, \ and\ \bibinfo {author}
  {\bibfnamefont {H.}~\bibnamefont {L{\"o}wen}},\ }\href {\doibase
  10.1126/sciadv.abf8998} {\bibfield  {journal} {\bibinfo  {journal} {Sci.
  Adv.}\ }\textbf {\bibinfo {volume} {7}},\ \bibinfo {pages} {eabf8998}
  (\bibinfo {year} {2021})}\BibitemShut {NoStop}%
\bibitem [{\citenamefont {Dauchot}\ and\ \citenamefont
  {D{\'e}mery}(2019)}]{dauchot2019}%
  \BibitemOpen
  \bibfield  {author} {\bibinfo {author} {\bibfnamefont {O.}~\bibnamefont
  {Dauchot}}\ and\ \bibinfo {author} {\bibfnamefont {V.}~\bibnamefont
  {D{\'e}mery}},\ }\href {\doibase
  https://doi.org/10.1103/PhysRevLett.122.068002} {\bibfield  {journal}
  {\bibinfo  {journal} {Phys. Rev. Lett.}\ }\textbf {\bibinfo {volume} {122}},\
  \bibinfo {pages} {068002} (\bibinfo {year} {2019})}\BibitemShut {NoStop}%
\bibitem [{\citenamefont {Engbring}\ \emph {et~al.}(2023)\citenamefont
  {Engbring}, \citenamefont {Boriskovsky}, \citenamefont {Roichman},\ and\
  \citenamefont {Lindner}}]{engbring2023}%
  \BibitemOpen
  \bibfield  {author} {\bibinfo {author} {\bibfnamefont {K.}~\bibnamefont
  {Engbring}}, \bibinfo {author} {\bibfnamefont {D.}~\bibnamefont
  {Boriskovsky}}, \bibinfo {author} {\bibfnamefont {Y.}~\bibnamefont
  {Roichman}}, \ and\ \bibinfo {author} {\bibfnamefont {B.}~\bibnamefont
  {Lindner}},\ }\href {\doibase https://doi.org/10.1103/PhysRevX.13.021034}
  {\bibfield  {journal} {\bibinfo  {journal} {Phys. Rev. X}\ }\textbf {\bibinfo
  {volume} {13}},\ \bibinfo {pages} {021034} (\bibinfo {year}
  {2023})}\BibitemShut {NoStop}%
\bibitem [{\citenamefont {Lepro}\ \emph {et~al.}(2022)\citenamefont {Lepro},
  \citenamefont {Gro{\ss}mann}, \citenamefont {Sharifi~Panah}, \citenamefont
  {Nagel}, \citenamefont {Klumpp}, \citenamefont {Lipowsky},\ and\
  \citenamefont {Beta}}]{lepro_optimal_2022}%
  \BibitemOpen
  \bibfield  {author} {\bibinfo {author} {\bibfnamefont {V.}~\bibnamefont
  {Lepro}}, \bibinfo {author} {\bibfnamefont {R.}~\bibnamefont {Gro{\ss}mann}},
  \bibinfo {author} {\bibfnamefont {S.}~\bibnamefont {Sharifi~Panah}}, \bibinfo
  {author} {\bibfnamefont {O.}~\bibnamefont {Nagel}}, \bibinfo {author}
  {\bibfnamefont {S.}~\bibnamefont {Klumpp}}, \bibinfo {author} {\bibfnamefont
  {R.}~\bibnamefont {Lipowsky}}, \ and\ \bibinfo {author} {\bibfnamefont
  {C.}~\bibnamefont {Beta}},\ }\href {\doibase
  10.1103/PhysRevApplied.18.034014} {\bibfield  {journal} {\bibinfo  {journal}
  {Phys. Rev. Appl.}\ }\textbf {\bibinfo {volume} {18}},\ \bibinfo {pages}
  {034014} (\bibinfo {year} {2022})}\BibitemShut {NoStop}%
\bibitem [{\citenamefont {Archer}\ \emph {et~al.}(2015)\citenamefont {Archer},
  \citenamefont {Campbell},\ and\ \citenamefont {Ebbens}}]{archer2015}%
  \BibitemOpen
  \bibfield  {author} {\bibinfo {author} {\bibfnamefont {R.}~\bibnamefont
  {Archer}}, \bibinfo {author} {\bibfnamefont {A.}~\bibnamefont {Campbell}}, \
  and\ \bibinfo {author} {\bibfnamefont {S.}~\bibnamefont {Ebbens}},\ }\href
  {\doibase 10.1039/C5SM01323B} {\bibfield  {journal} {\bibinfo  {journal}
  {Soft Matter}\ }\textbf {\bibinfo {volume} {11}},\ \bibinfo {pages} {6872}
  (\bibinfo {year} {2015})}\BibitemShut {NoStop}%
\end{thebibliography}%
\onecolumngrid
\appendix
\pagenumbering{Roman}
\renewcommand{\thefigure}{SI~\arabic{figure}}
\setcounter{figure}{0}
\renewcommand{\thetable}{SI~\arabic{table}}
\setcounter{table}{0}
\renewcommand{\theequation}{SI~\arabic{equation}}
\setcounter{equation}{0}
\vspace{.5in}
\begin{center}
    {\large \textbf{Supplemental Material: Active Transport of Cargo-Carrying and Interconnected Chiral Particles}}
\end{center}
\begin{center}
    Bhavesh Valecha,$^{1}$ Hossein Vahid,$^{2}$ Pietro Luigi Muzzeddu,$^{3}$ Jens-Uwe Sommer,$^{2,4}$ and Abhinav Sharma$^{1,2}$
\end{center}
\subsection{Outline}
\label{subsec:outline}
\noindent We present the detailed derivation of the steady state density distribution of the active-passive composite, for the cases where they are connected by an infinitely stiff spring and a spring with zero rest length. These notes are organised as follows: In Sections~\ref{subsec:coarsegraining1} and~\ref{subsec:coarsegraining2}, we perform the coarse graining of the Fokker-Planck equation by integrating out the rotational and bond degrees of freedom. Section~\ref{subsec:approximations} describes the approximations employed to arrive at the close form solution of the steady state density distribution. In Section~\ref{subsec:density}, we show the calculation for the tactic parameter for both the infinitely stiff spring and the spring with zero rest length. This is followed by simulation results on the tactic behavior of chains of chiral active particles in Section~\ref{subsec:chiral_chains}. We end with the details of the Langevin dynamics simulations in Section~\ref{subsec:simulation}.\vspace{0.2in}
\subsection{Eliminating the orientational degrees of freedom}
\label{subsec:coarsegraining1}
\noindent We begin with the stochastic equations of motion that govern the dynamics of the active-passive composite (Eq. (1) in the main text). These are given by:
\begin{subequations}
\begin{align}
    \frac{\dd\boldsymbol{ r}_1}{\dd t} &= \frac{1}{\gamma}\boldsymbol{F}+\frac{1}{\gamma}f_s(\boldsymbol{r}_1)\boldsymbol{p}+\sqrt{\frac{2T}{\gamma}}\,\boldsymbol{\xi}_1(t),\\
    \frac{\dd\theta}{\dd t} &= \omega + \sqrt{2D_R}\,\eta(t),\\
    \frac{\dd\boldsymbol{ r}_2}{\dd t} &= -\frac{1}{q\gamma}\boldsymbol{F}+\sqrt{\frac{2T}{q\gamma}}\,\boldsymbol{\xi}_2(t).
\end{align}
\label{eq:chiral_active_passive_dimer_dynamics_supplementary}
\end{subequations}   

\noindent It is convenient to make a coordinate transformation to the collective coordinate of the composite, which are identified as the center of friction and the bond coordinates, given by:
\begin{equation}
\begin{split}
    \boldsymbol{R}&=\frac{1}{1+q}\boldsymbol{r}_1+\frac{q}{1+q}\boldsymbol{r}_2,\\    
    \boldsymbol{r}&=\boldsymbol{r}_1-\boldsymbol{r}_2.
\end{split}
\label{eq:coordinate_transformation_supplementary}
\end{equation}
\noindent We now switch to an equivalent description in terms of the Fokker-Planck equation (FPE) for the probability density $P(\boldsymbol{R},\boldsymbol{r},\theta,t)\equiv P$, given by:

\begin{equation}
\begin{split}
    \dfrac{\partial}{\partial t}\,P =& -\nabla_{\boldsymbol{R}}\cdot\bigg[\dfrac{1}{1+q}\dfrac{1}{\gamma}f_{s}\boldsymbol{p}P - \dfrac{1}{1+q}\dfrac{T}{\gamma}\nabla_{\boldsymbol{R}}P\bigg] \\
    &-\nabla_{\boldsymbol{r}}\cdot\bigg[\dfrac{1+q}{q}\dfrac{1}{\gamma}\boldsymbol{F}P + \dfrac{1}{\gamma}f_{s}\boldsymbol{p}P -\dfrac{1+q}{q}\dfrac{T}{\gamma}\nabla_{\boldsymbol{r}}P\bigg]\\
    &-\omega\partial_{\theta}P+ D_R\partial_{\theta}^{2}P,   
\end{split}
\label{eq:fokker_planck_supplementary}
\end{equation}
where the symbol $\cdot$ represents a single contraction, and $\nabla_{\boldsymbol{R}}$ and $\nabla_{\boldsymbol{r}}$ represent derivatives with respect to $\boldsymbol{R}$ and $\boldsymbol{r}$, respectively. $\partial_{\theta}$ is the rotation operator in two dimensions and $\partial_{\theta}^{2}$ is the Laplacian on a unit circle.
\vspace{.2cm}

\noindent Following the analysis introduced in~\cite{vuijk_chemotaxis_2021}, we start by expanding the probability density in the eigenfunctions of $\partial_{\theta}^{2}$ operator:
\begin{equation}
  P(\boldsymbol{R},\boldsymbol{r},\theta,t) = \phi + \boldsymbol{\sigma}\cdot\boldsymbol{p} + \boldsymbol{\mu}\colon\boldsymbol{Q} + \boldsymbol{\Theta}(P).
\label{eq:multipole_expansion_supplementary}
\end{equation} 
Here, the $1$, $\boldsymbol{p}$ and $\boldsymbol{Q}=\boldsymbol{p}\,\boldsymbol{p}-\boldsymbol{\mathds{1}}/2$ are the first three eigenfunctions of $\partial_{\theta}^{2}$ operator with eigenvalues 0, -1 and -4 respectively, and $\colon$ denotes double contraction. Note that these eigenfunctions are equivalent to the $0^{\text{th}}$, $1^{\text{st}}$ and $2^{\text{nd}}$ order moments in the angular multipole expansion in two dimensions, and correspond to physically relevant observables. Specifically, $\phi(\boldsymbol{R},\boldsymbol{r},t)$ is the positional probability density,  $\boldsymbol{\sigma}(\boldsymbol{R},\boldsymbol{r},t)$ is proportional to the average orientation and  $\boldsymbol{\mu}(\boldsymbol{R},\boldsymbol{{r},t})$ is related to the nematic order parameter. $\boldsymbol{\Theta}(P)$ contains the dependencies on all the higher order moments.

\noindent Since we eventually want to know the collective position of the active-passive composite in the steady state, we will integrate out the orientational degrees of freedom. This forms our first coarse-graining step. To this end, we introduce the inner product:
\begin{equation}
\big\langle\,f(\boldsymbol{p}(\theta)),\,g(\boldsymbol{p}(\theta))\,\big\rangle = \int_{0}^{2\pi}\dd\theta f(\boldsymbol{p}(\theta))\,g(\boldsymbol{p}(\theta)).
\label{eq:inner_product}
\end{equation}

\noindent We can now project the FPE onto the eigenfunctions of $\partial_{\theta}^{2}$ operator to get the time evolution of the moments. To do this, we will need to use the following relations involving the orthogonality of the eigenfunctions of our expansion:
\begin{equation}
\begin{split}
    \langle\,1,1\,\rangle &= 2\pi, \\
    \langle\,\boldsymbol{p},1\,\rangle &=0,\\
    \langle\,\boldsymbol{Q},1\,\rangle &=0,\\
    \langle\,1,P\,\rangle&=2\pi \phi,\\
    \langle\,\boldsymbol{p}, P\,\rangle &=\pi\boldsymbol{\sigma},\\
    \langle\,\boldsymbol{Q},P\,\rangle &= \frac{\pi}{2}\boldsymbol{\mu},\\
    \langle\,\boldsymbol{p},\partial_{\theta}P\,\rangle &= -\pi\,\boldsymbol{\varepsilon}\cdot\boldsymbol{\sigma},\\   
    \langle\,1,\boldsymbol{\Theta}(P)\,\rangle&=0,\\
    \langle\,\boldsymbol{p}, \boldsymbol{\Theta}(P)\,\rangle&=0,\\
    \langle\,\boldsymbol{Q},\boldsymbol{\Theta}(P)\,\rangle &=0.\\
\end{split}
\label{eq:projection_relations}
\end{equation}

\noindent Note that, we have evaluated the inner product $\langle\,\boldsymbol{p},\partial_{\theta}P\,\rangle$ as follows: 
$$\langle\,\boldsymbol{p},\partial_{\theta}P\,\rangle = -\int_{0}^{2\pi}\dd\theta\big(\partial_{\theta}\boldsymbol{p}\big)P = -\int_{0}^{2\pi}\dd\theta\boldsymbol{\varepsilon}\cdot\boldsymbol{p}P = -\boldsymbol{\varepsilon}\cdot\langle\,\boldsymbol{p},P\,\rangle = -\pi\,\boldsymbol{\varepsilon}\cdot\boldsymbol{\sigma},$$
where we have used integration by parts and introduced the two dimensional \emph{Levi-Civita} tensor $\boldsymbol{\varepsilon}$:
$$\boldsymbol{\varepsilon} = \begin{pmatrix}
        0 & -1\\
        1 & 0
    \end{pmatrix}.$$
    
\noindent To get the equation of motion for the positional probability density $2\pi\phi(\boldsymbol{r}_1,\boldsymbol{r}_2,t)= \int_{0}^{2\pi}\dd\theta\,P(\boldsymbol{r}_1,\boldsymbol{r}_2,\boldsymbol{p},t)$, we project eq.~\eqref{eq:multipole_expansion_supplementary} onto the identity eigenfunction and get:
\begin{equation}
\begin{split}
    \frac{\partial}{\partial t}\,\phi = -\nabla_{\boldsymbol{R}}\cdot\bigg[-\frac{T}{\gamma}\frac{1}{1+q}\nabla_{\boldsymbol{R}}\phi + \frac{1}{1+q}\frac{1}{2\gamma}f_s\boldsymbol{\sigma}\bigg] -\nabla_{\boldsymbol{r}}\cdot\left[\frac{1}{2\gamma}f_s \boldsymbol{\sigma}+\frac{1+q}{q}\frac{1}{\gamma}\boldsymbol{F}\phi -\frac{T}{\gamma}\frac{1+q}{q}\nabla_{\boldsymbol{r}} \phi\right].
\end{split}
\label{eq:phi_dynamics_transformed_supplementary}
\end{equation}

Similarly, taking the inner product of eq.~\eqref{eq:multipole_expansion_supplementary} with $\boldsymbol{p}$ gives:
\begin{equation}
\begin{split}
    \bigg\{\frac{\partial}{\partial t}+\big(D_R\,\boldsymbol{\mathds{1}}-\omega\,\boldsymbol{\varepsilon}\big)\cdot\bigg\}\boldsymbol{\sigma} =& -\frac{1}{\gamma}\left(\frac{1}{1+q}\nabla_{\boldsymbol{R}} +\nabla_{\boldsymbol{r}} \right)(f_s\phi)-\frac{1}{\gamma}\frac{1+q}{q}\nabla_{\boldsymbol{r}}\cdot(\boldsymbol{F}\boldsymbol{\sigma})\\
    &+ \frac{T}{\gamma}\left(\frac{1}{1+q}\nabla_{\boldsymbol{R}}^{2} + \frac{1+q}{q}\nabla_{\boldsymbol{r}}^{2}\right)\boldsymbol{\sigma}
    -\frac{1}{2\gamma}\left(\frac{1}{1+q}\nabla_{\boldsymbol{R}}+\nabla_{\boldsymbol{r}}\right)\cdot(f_s\boldsymbol{\mu}).
\end{split}
\label{eq:sigma_dynamics_transformed_supplementary}
\end{equation}

Finally, we can consider the inner product with the nematic tensor $\boldsymbol{Q}$ to get:

\begin{equation}
\begin{split}
    \bigg\{\frac{\partial}{\partial t}+4D_R\bigg\}\boldsymbol{\mu} =& -\frac{1}{\gamma}\frac{1+q}{q}\nabla_{\boldsymbol{r}}\cdot(\boldsymbol{F}\boldsymbol{\mu}) + \frac{T}{\gamma}\left(\frac{1}{1+q}\nabla_{\boldsymbol{R}}^{2}+\frac{1+q}{q}\nabla_{\boldsymbol{r}}^{2}\right)\boldsymbol{\mu}\\
    & - \frac{2}{\pi\gamma}\bigg[\left(\frac{1}{1+q}\nabla_{\boldsymbol{R}}+\nabla_{\boldsymbol{r}}\right)f_{s}\boldsymbol{\sigma}\bigg]\colon\langle\,\boldsymbol{Q},\boldsymbol{Q}\,\rangle - \frac{2}{\pi}\omega\langle\boldsymbol{Q},\partial_{\theta}P\rangle\\
    & - \frac{2}{\pi\gamma}\left(\frac{1}{1+q}\nabla_{\boldsymbol{R}}+\nabla_{\boldsymbol{r}}\right)\cdot\langle f_s\boldsymbol{p}\boldsymbol{Q},\Theta(P)\rangle.
\label{eq:mu_dynamics_transformed_supplementary}
\end{split}
\end{equation}

\noindent The equations~\eqref{eq:phi_dynamics_transformed_supplementary}, \eqref{eq:sigma_dynamics_transformed_supplementary} and~\eqref{eq:mu_dynamics_transformed_supplementary}, together with the evolution equations for higher order expansion coefficients, form a hierarchical structure and exactly determine the full probability distribution $P(\boldsymbol{R,r},\theta,t)$. Unfortunately, such a hierarchy cannot be solved exactly. However, we will see in the following, this hierarchy can be closed by employing the small gradients approximation, which implies that all terms involving the nematic tensor are of order at least $\mathcal{O}(\nabla_{\boldsymbol{R}}^2)$ and thus can be ignored.

\subsection{Eliminating the bond coordinate}
\label{subsec:coarsegraining2}
\noindent  As we are interested in the marginal distribution of the position of the center of mass, we integrate out the information about the relative distance between the chiral active particle and the passive particle. This results in a continuity equation for the probability distribution $\rho(\boldsymbol{R}) = 2\pi\int \dd\boldsymbol{r}\phi(\boldsymbol{R,r},t)$ of the collective coordinate $\boldsymbol{R}$. This is done as follows:
\begin{equation}
\begin{split}
    \frac{\partial}{\partial t}\int \dd\boldsymbol{r}\phi(\boldsymbol{R,r},t) =&\,\, \frac{T}{\gamma}\frac{1}{1+q}\nabla_{\boldsymbol{R}}^{2}\int \dd\boldsymbol{r}\phi -\frac{1}{1+q}\frac{1}{2\gamma}\int \dd\boldsymbol{r}\nabla_{\boldsymbol{R}}\cdot(f_s\boldsymbol{\sigma}) \\
    &-\int \dd\boldsymbol{r}\bigg\{\nabla_{\boldsymbol{r}}\cdot\left[\frac{1}{2\gamma}f_s \boldsymbol{\sigma}+\frac{1+q}{q}\frac{1}{\gamma}\boldsymbol{F}\phi -\frac{T}{\gamma}\frac{1+q}{q}\nabla_{\boldsymbol{r}}\phi\right]\bigg\}.
\end{split}
\label{eq:continuity_calculation}
\end{equation}

\noindent Under periodic boundary conditions, the above equation reduces to the continuity equation:
\begin{equation}
    \frac{\partial}{\partial t}\rho = -\nabla_{\boldsymbol{R}}\cdot\boldsymbol{J},
\label{eq:continuity_equation_supplementary}
\end{equation}
where the probability current can be broken down into a part due to diffusion and a part due to the polarization:
\begin{equation}
    \boldsymbol{J} = \boldsymbol{J}_{D} + \boldsymbol{J}_{\boldsymbol{\sigma}},
\label{eq:prob_current_contributions}
\end{equation}
with 
\begin{equation}
    \boldsymbol{J}_{D} = -\frac{T}{\gamma}\frac{1}{1+q}\nabla_{\boldsymbol{R}}\rho,
\label{eq:diffusive_current}
\end{equation}
and
\begin{equation}
    \boldsymbol{J}_{\boldsymbol{\sigma}} = \frac{2\pi}{1+q}\frac{1}{2\gamma}\int\,\dd\boldsymbol{r}f_{s}\boldsymbol{\sigma}.
\label{eq:convective_current}
\end{equation}

\subsection{Approximations}
\label{subsec:approximations}
\noindent To find the steady state density distribution $\rho(\boldsymbol{R})$ from eq.~\eqref{eq:continuity_equation_supplementary}, we need to calculate $\boldsymbol{J_{\sigma}}$, which involves closing the hierarchy in eqs.~\eqref{eq:phi_dynamics_transformed_supplementary}, \eqref{eq:sigma_dynamics_transformed_supplementary} and~\eqref{eq:mu_dynamics_transformed_supplementary}. To do so, we employ two approximations that are central to this analysis. These are the adiabatic approximation and the small gradients approximation. 

\subsubsection{Adiabatic Approximation}
\noindent We identify the position probability distribution $\phi(\boldsymbol{R,r},t)$ as the slowest mode of our dynamics. Indeed, $\phi(\boldsymbol{R,r},t)$ satisfies a continuity equation, meaning that it is locally conserved and decays on timescales which are $\mathcal{O}(\nabla_{\boldsymbol{R}}^{-1})$. Additionally, all the higher order moments decay at much faster timescales due to the presence of sink-terms in their dynamics. Thus, they are effectively quasi-static at the timescales at which our quantity of interest $\phi(\boldsymbol{R,r},t)$ evolves. In particular, $\boldsymbol{\sigma}$ decays on a timescale given by the eigenvalues of the matrix $\big(D_R\,\boldsymbol{\mathds{1}}-\omega\,\boldsymbol{\varepsilon}\big)^{-1}\equiv\boldsymbol{\mathds{L}}$ and $\boldsymbol{\mu}$ decays on timescales of the order of $(4D_{R})^{-1}$. This allows us to set the time derivative terms in eq.~\eqref{eq:sigma_dynamics_transformed_supplementary} and eq.~\eqref{eq:mu_dynamics_transformed_supplementary} to zero, giving us:

\begin{equation}
\begin{split}
    \boldsymbol{\sigma} = \boldsymbol{\mathds{L}}\cdot\bigg\{&-\frac{1}{\gamma}\left(\frac{1}{1+q}\nabla_{\boldsymbol{R}} +\nabla_{\boldsymbol{r}}\right)(f_s\phi)-\frac{1}{\gamma}\frac{1+q}{q}\nabla_{\boldsymbol{r}}\cdot(\boldsymbol{F}\boldsymbol{\sigma})\\ 
    &+\frac{T}{\gamma}\left(\frac{1}{1+q}\nabla_{\boldsymbol{R}}^{2}+\frac{1+q}{q}\nabla_{\boldsymbol{r}}^{2}\right)\boldsymbol{\sigma}
    -\frac{1}{2\gamma}\left(\frac{1}{1+q}\nabla_{\boldsymbol{R}}+\nabla_{\boldsymbol{r}}\right)\cdot(f_s\boldsymbol{\mu})\bigg\}
\end{split}
\label{eq:sigma_adiabatic_supplementary}
\end{equation}
and
\begin{equation}
\begin{split}
    \boldsymbol{\mu} = \frac{1}{4D_r}\bigg\{&-\frac{1}{\gamma}\frac{1+q}{q}\nabla_{\boldsymbol{r}}\cdot(\boldsymbol{F}\boldsymbol{\mu}) +\frac{T}{\gamma}\left(\frac{1}{1+q}\nabla_{\boldsymbol{R}}^{2}+\frac{1+q}{q}\nabla_{\boldsymbol{r}}^{2}\right)\boldsymbol{\mu}\\
    &-\frac{2}{\pi\gamma}\bigg[\Big(\frac{1}{1+q}\nabla_{\boldsymbol{R}}+\nabla_{\boldsymbol{r}}\Big)f_{s}\boldsymbol{\sigma}\bigg]\colon\langle\,\boldsymbol{Q},\boldsymbol{Q}\,\rangle-\frac{2}{\pi}\omega\langle\,\boldsymbol{Q},\mathcal{\partial_{\theta}}P\,\rangle\\
    &-\frac{2}{\pi\gamma}\left(\frac{1}{1+q}\nabla_{\boldsymbol{R}}+\nabla_{\boldsymbol{r}}\right)\cdot\langle\,f_s\boldsymbol{p}\,\boldsymbol{Q},\Theta(P)\,\rangle\bigg\}.
\end{split}
\label{eq:mu_adiabatic}
\end{equation}

Plugging eq.~\eqref{eq:sigma_adiabatic_supplementary} back into eq.~\eqref{eq:convective_current}, we get:
\begin{equation}
\begin{split}
    \boldsymbol{J}_{\boldsymbol{\sigma}} = 2\pi\boldsymbol{\mathds{L}}\cdot\bigg\{&-\frac{1}{(1+q)^2}\frac{1}{2\gamma^2}\int \dd\boldsymbol{r}f_{s}\nabla_{\boldsymbol{R}}(f_{s}\phi)-\frac{1}{1+q}\frac{1}{2\gamma^2}\int \dd\boldsymbol{r}f_{s}\nabla_{\boldsymbol{r}}(f_{s}\phi) \\
    &-\frac{1}{q}\frac{1}{2\gamma^{2}}\int \dd\boldsymbol{r}f_{s}\nabla_{\boldsymbol{r}}\cdot(\boldsymbol{F\sigma})+\frac{1}{(1+q)}\frac{T}{2\gamma^2}\int \dd\boldsymbol{r}f_{s}\bigg[\frac{1}{(1+q)}\nabla_{\boldsymbol{R}}\cdot(\nabla_{\boldsymbol{R}}\boldsymbol{\sigma})+\frac{(1+q)}{q}\nabla_{\boldsymbol{r}}\cdot(\nabla_{\boldsymbol{r}}\boldsymbol{\sigma})\bigg]\\
    &-\frac{1}{1+q}\frac{1}{4\gamma^2}\int \dd\boldsymbol{r}f_{s}\left(\frac{1}{1+q}\nabla_{\boldsymbol{R}}+\nabla_{\boldsymbol{r}}\right)\cdot (f_{s}\boldsymbol{\mu})\bigg\}.
\end{split}
\label{eq:convective_current_sub}
\end{equation}

\subsubsection{Small Gradients Approximation}
\noindent Next, we are interested in the limits where the gradients in the activity field are small compared to the persistence length of the chiral active particle as well as small compared to the separation between the active and the passive particles.
This approximation will allow us to decouple eq.~\eqref{eq:continuity_equation_supplementary},~\eqref{eq:prob_current_contributions} and ~\eqref{eq:convective_current_sub} to get a closed form solution of the marginal probability density $\rho(\boldsymbol{R})$.\\

\noindent To apply this approximation, it is important to realize that, in the absence of gradients in the activity field, our system has no directional preferences and will be isotropic in space. This implies that the nematic tensor $\boldsymbol{\mu}$ and higher order moments are at least $\mathcal{O}(\nabla_{\boldsymbol{R}})$. This allows us to significantly simplify eq.~\eqref{eq:convective_current_sub}, wherein we only have to evaluate the first three terms. We will also convert the gradients in $\nabla_{\boldsymbol{r}}$ to gradients in $\nabla_{\boldsymbol{R}}$ by applying integration by parts:
\begin{equation}
    \nabla_{\boldsymbol{r}}f_{s}(\bm{r}_1) = \frac{q}{1+q}\nabla_{\boldsymbol{R}}f_{s}(\bm{r}_1).
\end{equation}
We then get the following expression for $\boldsymbol{J}_{\boldsymbol{\sigma}}$:
\begin{equation}
\begin{split}
    \boldsymbol{J}_{\boldsymbol{\sigma}} = 2\pi\boldsymbol{\mathds{L}}\cdot\bigg\{-\frac{1}{(1+q)^2}\frac{1}{2\gamma^2}\int \dd\boldsymbol{r}f_{s}\nabla_{\boldsymbol{R}}(f_{s}\phi)+\frac{q}{(1+q)^2}\frac{1}{2\gamma^2}\int \dd\boldsymbol{r}\phi f_{s}\nabla_{\boldsymbol{R}}f_{s} + \frac{1}{1+q}\frac{1}{2\gamma^{2}}\boldsymbol{I}\bigg\},
\end{split}
\label{eq:convective_current_sub2}
\end{equation}
\noindent where we have defined the quantity $\boldsymbol{I}$ as:
\begin{equation}
    \boldsymbol{I} = \int \dd\boldsymbol{r}(\boldsymbol{F}\cdot\nabla_{\boldsymbol{R}}f_{s})\boldsymbol{\sigma}.
\label{eq:recurrace_integral1}
\end{equation}
We solve this integral as follows:
\begin{align}
    \boldsymbol{I} &= \int \dd\boldsymbol{r}(\boldsymbol{F}\cdot\nabla_{\boldsymbol{R}}f_{s})\boldsymbol{\sigma} \nonumber\\
    &= -\frac{1}{\gamma}\boldsymbol{\mathds{L}}\cdot\int \dd\boldsymbol{r}\big[\boldsymbol{F}\cdot(\nabla_{\boldsymbol{R}}f_{s})\nabla_{\boldsymbol{r}}\big]\cdot\bigg(\boldsymbol{\mathds{1}}\phi f_{s} + \frac{1+q}{q}\boldsymbol{F\sigma}\bigg) + \mathcal{O}(\nabla_{\boldsymbol{R}}^{2} f_{s})\nonumber \\
    &= \frac{1}{\gamma}\boldsymbol{\mathds{L}}\cdot\int \dd\boldsymbol{r} \nabla_{\boldsymbol{r}}(\boldsymbol{F}\cdot\nabla_{\boldsymbol{R}}f_{s}) \cdot \bigg(\boldsymbol{\mathds{1}}\phi f_{s} +\frac{1+q}{q}\boldsymbol{F\sigma}\bigg) + \mathcal{O}(\nabla_{\boldsymbol{R}}^{2}f_{s}),\label{eq:recurrace_integral2}
\end{align}
where we have used integration by parts in the last line. Now, we use
\begin{equation}
    \nabla_{\boldsymbol{r}}(\boldsymbol{F}\cdot\nabla_{\boldsymbol{R}}f_{s}) = \nabla_{\boldsymbol{r}}\boldsymbol{F}\cdot\nabla_{\boldsymbol{R}}f_{s} + \mathcal{O}(\nabla_{\boldsymbol{R}}^{2}f_s). \nonumber
\end{equation}
The interaction between the chiral active particle and the passive particle is modeled as a harmonic spring with rest length $l_0$. So we can evaluate $\nabla_{\boldsymbol{r}}\boldsymbol{F}$ as:
\begin{equation}
    \nabla_{\boldsymbol{r}}\boldsymbol{F} = -k\nabla\boldsymbol{r}[(r-l_0)\boldsymbol{\hat{r}}] = -k\big[\boldsymbol{\hat{r}\hat{r}} + (1 - l_0/r)(\boldsymbol{1}-\boldsymbol{\hat{r}\hat{r}})\big]\equiv -k\boldsymbol{\mathds{A}},\,\,\text{where}\,\,r = |\boldsymbol{r}|. \nonumber
\end{equation}
With this, $\boldsymbol{I}$ becomes
\begin{align}
    \boldsymbol{I} &= -\frac{k}{\gamma}\boldsymbol{\mathds{L}}\cdot\bigg[\int \dd\boldsymbol{r}\phi f_{s}\boldsymbol{\mathds{A}}\cdot\nabla_{\boldsymbol{R}}f_s + \frac{1+q}{q}\int \dd\boldsymbol{r}(\boldsymbol{F}\cdot\nabla_{\boldsymbol{R}}f_{s})\boldsymbol{\sigma}\bigg] \nonumber \\
    &= -\frac{k}{\gamma}\boldsymbol{\mathds{L}}\cdot\bigg[\int \dd\boldsymbol{r}\phi f_{s}\boldsymbol{\mathds{A}}\cdot\nabla_{\boldsymbol{R}}f_s + \frac{1+q}{q}\boldsymbol{I}\bigg] \nonumber \\
    &= -\dfrac{qk}{\gamma}\boldsymbol{\mathds{L}}\cdot\bigg(q\boldsymbol{\mathds{1}}+\dfrac{(1+q)k}{\gamma}\boldsymbol{\mathds{L}}\bigg)^{-1}\int \dd\boldsymbol{r}\phi f_{s}\boldsymbol{\mathds{A}}\cdot\nabla_{\boldsymbol{R}}f_s \nonumber \\
    &= -\boldsymbol{\mathds{B}}\cdot\int\dd\boldsymbol{r}\phi f_{s}\boldsymbol{\mathds{A}}\cdot\nabla_{\boldsymbol{R}}f_s.\label{eq:recurrance_integral_final}
\end{align}
\noindent where we used the fact that the matrix $\mathds{A}$ is symmetric, the identity $\mathds{A}\cdot \bm{F}=\bm{F}$ and we have simplified the notation by introducing the quantity $\boldsymbol{\mathds{B}}$:
$$\boldsymbol{\mathds{B}} = \dfrac{qk}{\gamma}\boldsymbol{\mathds{L}}\bigg(q\boldsymbol{\mathds{1}}+\dfrac{(1+q)k}{\gamma}\boldsymbol{\mathds{L}}\bigg)^{-1}.$$\\

\noindent We can now finally write the closed form expression for the polarisation current $\boldsymbol{J}_{\boldsymbol{\sigma}}$
\begin{equation}
    \boldsymbol{J}_{\boldsymbol{\sigma}} = -\frac{2\pi}{(1+q)^2}\frac{1}{2\gamma^2}\boldsymbol{\mathds{L}}\cdot\bigg[\int \dd\boldsymbol{r}f_{s}\nabla_{\boldsymbol{R}}(f_{s}\phi) - q\int \dd\boldsymbol{r}\phi f_{s}\nabla_{\boldsymbol{R}}f_{s} + (1+q)\boldsymbol{\mathds{B}}\int \dd\boldsymbol{r}\phi f_{s}\boldsymbol{\mathds{A}}\cdot\nabla_{\boldsymbol{R}}f_{s}\bigg].
\end{equation}

\noindent Now, we are in the position to get the expression of the steady state density distribution $\rho$. In particular, we analyze two specific forms of the force $\boldsymbol{F}$ between the active particle and the passive particle: the rigid bond case and the harmonic potential with zero rest length. We do this in the following section.

\subsection{Steady state density distribution}
\label{subsec:density}
\subsubsection{Spring with zero rest length}
\noindent In this case, the quantity $\boldsymbol{\mathds{A}}$ defined for the spring force is the identity matrix, i.e., $\boldsymbol{\mathds{A}}(l_0=0) = \boldsymbol{\mathds{1}}$. Since the spring has zero rest length, the separation between the active and passive particles is small compared to the gradients of the activity field, provided that $k$ is not too small. So, we can approximate $\phi(\boldsymbol{R,r},t)$ as:

\begin{equation}
\phi(\boldsymbol{R,r},t)\approx\frac{1}{2\pi}\rho(\boldsymbol{R})\delta(\boldsymbol{r}).
\label{eq:phi_spring_zero_rest_length}
\end{equation}
\noindent Moreover, we can also Taylor-expand the activity field about the collective coordinate $\boldsymbol{R}$ as:
\begin{equation}
    f_{s} = f_{s}(\boldsymbol{r}_1) = f_{s}\left(\boldsymbol{R} + \boldsymbol{r}\frac{q}{1+q}\right) = f_{s}(\boldsymbol{R}) + \frac{q}{1+q}\bm{r}\cdot \nabla_{\boldsymbol{R}}f_{s}(\bm{R}),
\label{eq:activity_taylor_expansion}
\end{equation}
\noindent and thus move it out of the $\boldsymbol{r}$ integral in all terms.\\

\noindent With these steps, we can write $\boldsymbol{J}_{\boldsymbol{\sigma}}$ as:

\begin{align}
    \boldsymbol{J}_{\boldsymbol{\sigma}} =& -\frac{2\pi}{(1+q)^2}\frac{1}{2\gamma^2}\boldsymbol{\mathds{L}}\cdot\bigg[f_{s}^{2}\nabla_{\boldsymbol{R}}\rho\int d\boldsymbol{r}\frac{1}{2\pi}\delta(\boldsymbol{r}) + \rho f_{s}\nabla_{\boldsymbol{R}}f_{s}\int d\boldsymbol{r}\frac{1}{2\pi }\delta(\boldsymbol{r})\nonumber\\
    &-q\rho f_s\nabla_{\boldsymbol{R}}f_s\int d\boldsymbol{r}\frac{1}{2\pi}\delta(\boldsymbol{r}) +(1+q)\boldsymbol{\mathds{B}}f_s\nabla_{\boldsymbol{R}}f_s\rho\int d\boldsymbol{r}\frac{1}{2\pi}\delta(\boldsymbol{r})\bigg] + \mathcal{O}(\nabla_{\boldsymbol{R}}^{2})\nonumber\\
    =& -\frac{1}{(1+q)^2}\frac{1}{2\gamma^2}f_{s}^{2}\boldsymbol{\mathds{L}}\cdot\nabla_{\boldsymbol{R}}\rho - \frac{1}{(1+q)^2}\frac{1}{2\gamma^2}\frac{\rho}{2}\boldsymbol{\mathds{L}}\big[(1-q)\boldsymbol{\mathds{1}}+(1+q)\boldsymbol{\mathds{B}}\big]\cdot\nabla_{\boldsymbol{R}}(f_{s}^{2}) +\mathcal{O}(\nabla_{\boldsymbol{R}}^{2}),
\end{align}
We can now evaluate the total flux up to the drift/diffusion order as:
\begin{align}
    \boldsymbol{J} =& \,\boldsymbol{J}_D + \boldsymbol{J}_{\boldsymbol{\sigma}}\nonumber\\
    =&-\frac{1}{(1+q)}\frac{T}{\gamma}\nabla_{\boldsymbol{R}}\rho - \frac{1}{(1+q)^2}\frac{1}{2\gamma^2}f_{s}^{2}\boldsymbol{\mathds{L}}\cdot\nabla_{\boldsymbol{R}}\rho - \frac{1}{(1+q)^2}\frac{1}{2\gamma^2}\frac{\rho}{2}\boldsymbol{\mathds{L}}\big[(1-q)\boldsymbol{\mathds{1}}+(1+q)\boldsymbol{\mathds{B}}\big]\cdot\nabla_{\boldsymbol{R}}(f_{s}^{2}).
\end{align}
In particular, the total flux $\bm{J}$ has the following structure
\begin{equation}
    \boldsymbol{J} = \boldsymbol{V}(\boldsymbol{R})\rho(\boldsymbol{R}) - \nabla_{ \boldsymbol{R}}  \cdot \left(\boldsymbol{\mathds{D}}(\boldsymbol{R})\rho(\boldsymbol{R})\right),
\label{eq:drift_diffusion_equation_supplementary1}
\end{equation}
where the effective diffusion coefficient depends on $\boldsymbol{R}$ and is given by:
\begin{equation}
  \boldsymbol{\mathds{D}}(\boldsymbol{R}) = \dfrac{1}{1+q}\frac{T}{\gamma}\boldsymbol{\mathds{1}} + \dfrac{1}{(1+q)^2}\dfrac{1}{2\gamma^2}f_{s}^{2}(\boldsymbol{R})\boldsymbol{\mathds{L}}^{T},
\label{eq:effective_diffusion_supplementary}  
\end{equation}  
and the effective drift $\boldsymbol{V}(\boldsymbol{R})$ can be written in terms of $\boldsymbol{\mathds{D}}(\boldsymbol{R})$ as:
\begin{equation}
\begin{split}
  \boldsymbol{V}(\boldsymbol{R}) = \left(\boldsymbol{\mathds{1}}-\frac{1}{2}  \boldsymbol{\mathds{L}}\big[(1-q)\boldsymbol{\mathds{1}}+(1+q)\boldsymbol{\mathds{B}}\big]\boldsymbol{\mathds{L}}^{-1}\right)\nabla_{\boldsymbol{R}}\cdot \boldsymbol{\mathds{D}}(\boldsymbol{R}).
\label{eq:effective_drift_supplentary}
\end{split}
\end{equation}

\noindent In this study, the activity gradients are assumed to exist only in the $x$-direction. Since there is translational invariance in the $y$ direction, the stationary density $\rho$ has variations only along the $x$-direction. Thus, the stationary probability flux along the $x$-direction is given by:
\begin{equation}
\begin{split}
\label{eq:x_flux_supplementary}
    J_x(x) =& -\frac{1}{(1+q)}\frac{T}{\gamma}\partial_{x}\rho(x) - \frac{1}{(1+q)^2}\frac{1}{2\gamma^2}f_{s}^{2}(x)\boldsymbol{\mathds{L}}_{xx}\partial_{x}\rho(x)\\
    &- \frac{1}{(1+q)^2}\frac{1}{2\gamma^2}\frac{\rho(x)}{2}\bigg\{\boldsymbol{\mathds{L}}\big[(1-q)\boldsymbol{\mathds{1}}+(1+q)\boldsymbol{\mathds{B}}\big]\bigg\}_{xx}\partial_{x}(f_{s}(x)^{2})\\
    =&-\frac{\epsilon}{2} \rho \partial_x \mathds{D}_{xx} - \mathds{D}_{xx} \partial_x \rho,
\end{split}
\end{equation}
where we denoted with $\mathds{D}_{xx}$ the $xx$ element of the effective diffusion coefficient.
We define the quantity $\epsilon$ as the \emph{tactic parameter}, which determines the accumulation behavior of the active-passive composite. Particularly, for $\epsilon<0$, the composite accumulates in regions of high activity and vice-versa. Imposing a zero-flux condition along the direction of the activity gradient, we obtain the steady state density:
\begin{equation}
  \rho(x) \propto \bigg[1+\dfrac{D_R}{D_{R}^{2}+\omega^2}\dfrac{1}{2\gamma T}\dfrac{1}{(1+q)}f_{s}^{2}(x)\bigg]^{-\displaystyle \epsilon/2}, 
\label{eq:steady_state_density_supplementary}
\end{equation}
where the tactic parameter enters as the exponent, and is dependent on the chiral torque $\Omega$. Specifically, we get
\begin{equation}
\begin{split}
  \epsilon = 1 - q\dfrac{\big(1+\Omega^2\big)\big(1+\tau\big)}{\Omega^2 + (1+\tau)^2},
\end{split}  
  \label{eq:epsilon_spring_supplementary}
\end{equation}  
where we have introduced the following non-dimensional parameters in units of the persistence time $\tau_p = D_{R}^{-1}$ of the active particle due to rotational diffusion,
$$\Omega=\omega\tau_p\quad\text{and}\quad\tau = \frac{(1+q)k\tau_p}{q\gamma}.$$
%$$\Omega=\frac{\omega}{D_{R}}\quad\text{and}\quad\tau = \frac{(1+q)k}{q\gamma D_{R}}.$$

\subsubsection{Infinitely stiff spring}
\noindent This is implemented by taking the limit of the spring constant $k\rightarrow\infty$. The only term where the spring constant appears is $\boldsymbol{\mathds{B}}$ and we can now evaluate it in this limit as:
\begin{equation}
    \lim_{k\rightarrow\infty}(1+q)\boldsymbol{\mathds{B}} = \lim_{k\rightarrow\infty}\dfrac{q(1+q)k}{\gamma}\boldsymbol{\mathds{L}}\bigg(q\boldsymbol{\mathds{1}}+\dfrac{(1+q)k}{\gamma}\boldsymbol{\mathds{L}}\bigg)^{-1} = q\boldsymbol{\mathds{1}}.
\label{eq:B_stiff_spring}
\end{equation}
\noindent Note that in this limit, the spring force between the chiral active particle and the passive particle is so strong that the separation between them approximately remains the same as the spring rest length $l_0$, i.e., $r'\approx l_0$ and so we can write: 
\begin{equation}
    \phi(\boldsymbol{R},\boldsymbol{r}',t)\approx \frac{1}{2\pi} \frac{1}{2\pi l_0}\rho (\boldsymbol{R},t)\delta(r'-l_0).
    \label{eq:phi_stiff_spring}
\end{equation}
\noindent Additionally, we can also Taylor-expand the activity field in the same way as Eq.~\eqref{eq:activity_taylor_expansion}. With these steps, $\boldsymbol{J_{\sigma}}$ now becomes
\begin{align}
    \boldsymbol{J}_{\boldsymbol{\sigma}} =& -\frac{1}{(1+q)^2}\frac{1}{2\gamma^2}\boldsymbol{\mathds{L}}\cdot\bigg[f_{s}^{2}\nabla_{\boldsymbol{R}}\rho\int \dd\boldsymbol{r}\frac{1}{2\pi l_0}\delta(r-l_0) + \rho f_{s}\nabla_{\boldsymbol{R}}f_{s}\int \dd\boldsymbol{r}\frac{1}{2\pi l_0}\delta(r-l_0)\nonumber\\
    &-q\rho f_s\nabla_{\boldsymbol{R}}f_s\int \dd\boldsymbol{r}\frac{1}{2\pi l_0}\delta(r-l_0) +q\,\boldsymbol{\mathds{1}}f_s\nabla_{\boldsymbol{R}}f_s\rho\int \dd\boldsymbol{r}\frac{1}{2\pi l_0}\delta(r-l_0)\boldsymbol{\hat{r}\hat{r}}\bigg]+\mathcal{O}(\nabla_{\boldsymbol{R}^{2}})\nonumber \\
    =& -\frac{1}{(1+q)^2}\frac{1}{2\gamma^2}\boldsymbol{\mathds{L}}\cdot\bigg[f_{s}^{2}\nabla_{\boldsymbol{R}}\rho+\frac{\rho}{2}\nabla_{\boldsymbol{R}}(f_{s}^{2})-\frac{q\rho}{2}\nabla_{\boldsymbol{R}}(f_{s}^{2})+\frac{q\rho}{4}\nabla_{\boldsymbol{R}}(f_{s}^{2})\bigg] +\mathcal{O}(\nabla_{\boldsymbol{R}^{2}})
    \label{eq:convective_current_stiff},
\end{align}
where we have used the normalization of the delta distribution and 
\begin{equation}
    \frac{1}{2\pi}\int \dd\boldsymbol{r}\boldsymbol{\hat{r}\hat{r}} = \frac{\boldsymbol{\mathds{1}}}{2}.
\end{equation}

\noindent The total flux then becomes:
\begin{align} 
    \boldsymbol{J} =& \boldsymbol{J}_{D} + \boldsymbol{J}_{\boldsymbol{\sigma}}, \nonumber\\
    =& -\frac{1}{(1+q)}\frac{T}{\gamma}\nabla_{\boldsymbol{R}}\rho - \frac{1}{(1+q)^2}\frac{1}{2\gamma^2}f_{s}^{2}\boldsymbol{\mathds{L}}\cdot\nabla_{\boldsymbol{R}}\rho\nonumber\\
    &-\left(1-\frac{q}{2}\right)\frac{1}{(1+q)^2}\frac{1}{2\gamma^2}\frac{\rho}{2}\boldsymbol{\mathds{L}}\cdot\nabla_{\boldsymbol{R}}(f_{s}^{2}).
\end{align}
Using this flux, we can write an effective drift-diffusion equation:
\begin{equation}
    \boldsymbol{J} = \boldsymbol{V}(\boldsymbol{R})\rho(\boldsymbol{R}) - \nabla_{ \boldsymbol{R}}  \cdot \left(\boldsymbol{\mathds{D}}(\boldsymbol{R})\rho(\boldsymbol{R})\right)\nonumber,
\label{eq:drift_diffusion_equation_supplementary2}
\end{equation}
We again consider activity to be varying only along the $x$-direction with the stationary flux given by:
\begin{equation}
\begin{split}
\label{eq:density_stiff_spring_supplementary}
     J_x(x) =& -\frac{1}{(1+q)}\frac{T}{\gamma}\partial_{x}\rho(x) - \frac{1}{(1+q)^2}\frac{1}{2\gamma^2}\frac{D_R}{D_{R}^{2}+\omega^2}f_{s}^{2}(x)\partial_{x}\rho(x)\\
     &-\frac{1}{2}\left(1-\frac{q}{2}\right)\frac{1}{(1+q)^2}\frac{1}{2\gamma^2}\frac{D_R}{D_{R}^{2}+\omega^2}\partial_{x}(f_{s}^{2}(x))\\
     =&-\frac{\epsilon}{2} \rho \partial_x \mathds{D}_{xx} - \mathds{D}_{xx} \partial_x \rho,
\end{split}    
\end{equation}
Similar to the previous case, we find that the effective drift and diffusion coefficient are related by a derivative relation. 
Upon imposing a zero-flux condition, we obtain the same structure for the steady state density:
\begin{equation}
  \rho(x) \propto \bigg[1+\dfrac{D_R}{D_{R}^{2}+\omega^2}\dfrac{1}{2\gamma T}\dfrac{1}{(1+q)}f_{s}^{2}(x)\bigg]^{-\displaystyle \epsilon/2}, 
\label{eq:steady_state_density_stiff_supplementary}
\end{equation}
However, the tactic parameter $\epsilon$ is now given by:
\begin{equation}
  \epsilon = 1 - \frac{q}{2}.
\label{eq:epsilon_stiff_supplementary}
\end{equation}
We see that the tactic parameter $\epsilon$ is only dependent on the size of the passive particle via $q$.

\subsection{Chiral Active Chains}
\label{subsec:chiral_chains}
\begin{figure}[H]
  \centering
  \includegraphics[width=0.9\linewidth]{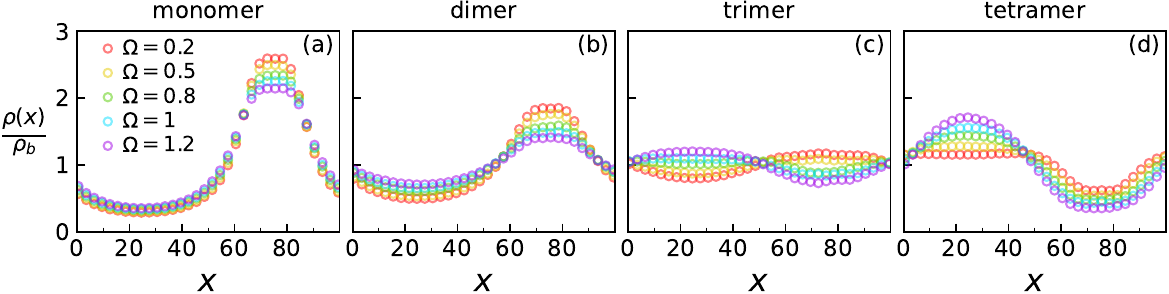}
    \caption{The steady state density distribution of active (a) monomers, (b) dimers, (c) trimers, and (d) tetramers connected via a spring with $\ell_0=1.5$ and $k=30$. The $y$-axis is normalized by the bulk density $\rho_b$. The sinusoidal activity field is the same as in Fig. (2) of main manuscript, given by $f_s(x) = 20 \left(1 + \sin\left({2\pi x}/{L}\right)\right)$. 
    The accumulation behavior of the chains is influenced by the strength of the chiral torque $\Omega$. The parameters of the simulation are $k_{\rm B} T= 1.0$, $\gamma = 1.0$, $D_{R} = 10.0$, and the integration time step $\Delta t = 5D_R \times 10^{-6}$.}
    \label{fig:6}
\end{figure}

In the case of monomers and chains of chiral active particles without passive particles attached, we observe a dependence of chemotactic behavior on the chiral torque $\Omega$. Figure~\ref{fig:6} shows as $\Omega$ increases, monomers and dimers exhibit a reduction in antichemotaxis, but this effect does not vanish entirely. However, for trimers, a transition occurs at sufficiently high $\Omega$, where chains accumulate in the high-activity regions. This trend becomes even more pronounced in tetramers, where even low-chirality chains have a tendency to accumulate in high-activity regions.

\subsection{Simulation Details}
\label{subsec:simulation}
\noindent The chiral active particle and the passive particle were simulated using Langevin dynamics: The equations of motion (eq. (1) in the main text) were first discretized up to linear order in the integration time-step using the Euler method. The increments at each time-step were then summed up using the It\'o rule with a time-step size of $\dd t=D_R\times 10^{-5}$ for the case of harmonic spring with zero rest length and $\dd t=5D_R\times10^{-6}$ for the case of the infinitely stiff spring. The simulation box size was chosen to be $L=100$ with periodic boundary conditions. The activity field was also chosen to be periodic with sinusoidal variations.
%\bibliographystyle{vancouver}
%\bibliography{references}
\end{document}